\documentclass[10pt]{article}
\usepackage{epsfig}
\usepackage{amssymb,amsfonts}
\usepackage{amsmath}
\usepackage{multirow}
\usepackage{amsthm}
\usepackage{setspace}
\newcommand{\eps}{\varepsilon}

\newcommand{\Real}{\mathbb{R}}

 \newcommand{\Complex}{\mathbb{C}}

\def \prob {\mathbb{P}}
\def \supp {\mbox{Supp}}
\def \sgn {\mbox{Sgn}}

\def \bg{\mathbf{g}}

\def \bH{\mathbf{H}}
\def \bW{\mathbf{W}}
\def \bG{\mathbf{G}}
\def \bW{\mathbf{W}}
\def \bV{\mathbf{V}}
\def \bLambda{\mathbf{\Lambda}}
\def \bv{\mathbf{v}}
\def \bX{\mathbf{X}}
\def \bD{\mathbf{D}}
\def \bU{\mathbf{U}}
\def \bV{\mathbf{V}}
\def \bZ{\mathbf{Z}}
\def \bS{\mathbf{S}}

\def \ex {\mathbf{E}}
\def \bN{\mathbf{N}}
\def \bW{\mathbf{W}}
\def \bY{\mathbf{Y}}
\def \bZ{\mathbf{Z}}
\def \bPi{\mathbf{\Pi}}
\def \bsg{\mathbf{\Sigma}}

\def \ind {\mathbb{I}}
\def \bI {\mathbf{I}}
\def \snr {\textit{SNR}}
\def \om {\omega}

\newcounter{arabiccount}
\def\barabiclist{\begin{list}{\arabic{arabiccount}.}{\usecounter{arabiccount}
        \setlength{\itemsep}{0pt}\setlength{\parsep}{0pt}
        \setlength{\partopsep}{0pt}\setlength{\topsep}{5pt}}}
\def\earabiclist{\end{list}}

\begin{document}
\title{Information theoretic bounds for Compressed Sensing}
\author{Shuchin~Aeron, Venkatesh~Saligrama and Manqi Zhao%\\
%\{shuchin, mqzhao, srv\} @bu.edu%
%\thanks{}% <-this % stops a space
\thanks{The authors are with the department of Electrical and Computer Engineering at
Boston University, MA -02215. They can be reached at \{shuchin, srv, mqzhao\}@bu.edu. This research was supported by the Presidential Early Career Award (PECASE)
N00014-02-100362, NSF CAREER award ECS 0449194. Venkatesh Saligrama was also supported by the MIT-Portugal Program while he was visiting MIT.}}

%\markboth{IEEE Trans. on Info. Theory,~Vol. ?, No.? ,~November~
%?}{Shell \MakeLowercase{\textit{et al.}}: Bare Demo of IEEEtran.cls
%for Journals}
\date{}
\maketitle
\newtheorem{condition}{\indent \bf Condition}[section]
\newtheorem{property}{\indent \bf Property}[section]
\newtheorem{defn}{\indent \bf Definition}[section]
\newtheorem{conj}{\indent \bf Conjecture}[section]
\newtheorem{cor}{\indent \bf Corollary}[section]
\newtheorem{lem}{\indent \bf Lemma}[section]
\newtheorem{claim}{\indent \bf Claim}[section]
\newtheorem{thm}{\indent \bf Theorem}[section]
\newtheorem{prop}{\indent \bf Proposition}[section]
\newtheorem{remark}{\indent \bf Remark}[section]
\newtheorem{example}{\indent \bf Example}[section]

\begin{abstract}
In this paper we derive information theoretic performance bounds to sensing and reconstruction of sparse phenomena from noisy projections. We consider two settings: output noise
models where the noise enters after the projection and input noise models where the noise enters before the projection. We consider two types of distortion for reconstruction: support errors and mean-squared errors. Our goal is to relate the number of measurements, $m$, and $\snr$, to signal sparsity, $k$, distortion level, $d$, and signal dimension, $n$.

We consider support errors in a worst-case setting. We
employ different variations of Fano's inequality to derive necessary conditions on the number of measurements and $\snr$ required for exact reconstruction. To derive sufficient conditions we develop new insights on max-likelihood analysis based on a novel superposition property. In
particular this property implies that small support errors are the dominant error events. Consequently, our ML analysis does not suffer the conservatism of the
union bound and leads to a tighter analysis of max-likelihood. These results provide order-wise tight bounds. For output noise models we show that asymptotically an $\snr$ of $\Theta(\log(n))$ together with $\Theta(k \log(n/k))$ measurements is necessary and sufficient for exact support recovery. Furthermore, if a
small fraction of support errors can be tolerated, a constant $\snr$ turns out to be sufficient in the linear sparsity regime. In contrast for input noise models we show that support recovery fails if the
number of measurements scales as $o(n\log(n)/SNR)$ implying poor compression performance for such cases.

Motivated by the fact that the worst-case setup requires significantly high $\snr$ and substantial number of measurements for input and output noise models, we consider a Bayesian setup. To derive necessary conditions we develop novel extensions to Fano's inequality to handle continuous domains and arbitrary distortions. We then develop a new max-likelihood analysis over the set of rate distortion quantization points to characterize tradeoffs between mean-squared distortion and the number of measurements using rate-distortion theory. We show that with constant $\snr$ the number of measurements scales linearly with the rate-distortion function of the sparse phenomena.
\end{abstract}

\section{Introduction}
In this paper we derive information theoretic bounds on the performance of the Compressed Sensing problem,
\cite{Donoho1},\cite{CandesIT05},\cite{CandesIT06},
\begin{align}
\label{eq:model_snr} \bY = \bG \bX + \frac{\bN}{\sqrt{\snr}}
\end{align}
where the measurements $\bY \in \Real^{m\times 1}$, the desired signal $\bX \in \Real^n$, and the compression (sensing) matrix $\bG \in \Real^{m \times n}$. The noise $\bN \stackrel{d}{\sim} {\cal N}(0,\mathbf{I}_{m})$, where $\bI_m$ is an identity matrix of size $m$, is assumed to be a Gaussian
random vector with independent identically distributed (IID) components. We characterize results for both deterministic and stochastic compression matrices $\bG =[g_{ij}]$. For deterministic, $\bG$, the columns, $\bg_j$, are normalized to
have unit $\ell_2$ norm. For the stochastic setting we consider matrices drawn from IID (independent identically distributed) Gaussian ensembles. Each component here is assumed to be distributed as $g_{ij}
\stackrel{d}{\sim} {\cal N}(0,1/m),\,\,\,i=1,\,\ldots,m,\, j = 1,2,\ldots,\,n$. Note that under this normalized sensing matrix scenario, the term $\snr$ also denotes the inverse of the noise
variance. We refer to the signal model of Equation~(\ref{eq:model_snr}) as the \emph{output noise} model.  In parallel we also consider the \emph{input noise} model given by,
\begin{align}
\label{eq:model_snr1} \bY = \bG \left (\bX + {\bN \over \sqrt{\snr}} \right )
\end{align}
where $\bN \stackrel{d}{\sim} {\cal N}(0,\mathbf{I}_{n})$ is a Gaussian random vector with IID components. Evidently the noise here enters before
the ``compression'' operator, $\bG$, is applied. This model  is motivated by fusion problems that arise in sensor networks~\cite{NowakIT}, where noisy observations are
compressed.

The support of the signal $\bX$ is denoted by $\supp(\bX)=\{j \mid X_j \not = 0\}$. We assume that the cardinality of the signal support, $|\supp(\bX)| \leq k < n$.
It is often convenient to state and interpret results in terms of the sparsity ratio $\alpha_n = \frac{k}{n}$. The regime when $\alpha_n \stackrel{n \rightarrow
\infty}{\longrightarrow} \alpha>0$ is referred to as the linear regime and the regime when $\alpha_n \stackrel{n \rightarrow \infty}{\longrightarrow} 0$ is referred to as the
sub-linear regime.

We consider two types of distortions in signal reconstruction, namely, (a) Support distortion, i.e., $d(\hat{ \bX},\bX)) = \frac{1}{k}\sum_{j=1}^n |I_{\{X_i\not=0\}}-I_{\{\hat X_i\not=0\}}|$, where, $I_{\{\cdot\}}$ is the indicator function. (b)  Mean squared distortion, $d(\hat{
\bX},\bX)) = \frac{1}{n} \|\hat \bX - \bX\|^2 = \frac{1}{n} \sum_{j=1}^n |\hat X_j - X_j|^2$. These two distortion metrics address two different issues in signal recovery. The first metric penalizes solely the support detection part while the second metric penalizes both support detection and
amplitude estimation. We will now highlight the main contributions and results of the paper.

\subsection{Bounds for Exact and Approximate Support Recovery}
In this part we further restrict the signal $\bX$ to be bounded away from zero by a \emph{constant} $\beta > 0$ on its support. This is a standard assumption employed by other
researchers(see \cite{SrvZhao_arxiv,WainwrightITDec09,Candes07,WainwrightITMay09}) since it is impossible to identify the support of a signal $\bX$ from noisy measurements with arbitrarily small non-zero components. We derive necessary and sufficient conditions for exact and approximate support recovery
for this case under both output and input noise models. A central contribution of our work in this setting is that we explicitly quantify the required $\snr$ and the number of measurements, $m$ for exact support recovery. For the output noise model we show that the minimum $\snr$ required for support recovery is $\Omega(\log(n))$
regardless of $m$. In addition for this minimum $\snr$ level, the number of measurements, $m$, must scale as $\Omega(k\log(n/k))$ to guarantee exact support recovery.
Furthermore, we derive sufficient conditions and show that with $\snr=\Omega(\log(n))$ and $m=\Omega(k\log(n/k))$ the maximum-likelihood decoder can exactly identify the signal
support with high probability. These results are depicted in Table~\ref{tab:LASSOvsML}. While not depicted in this table it is interesting to consider what happens as $\snr$
increases.  The bounds derived in this paper show that we cannot get significant improvement in $m$ unless $\snr$ is scaled substantially (as a fractional power of $n$). We also
derive conditions for support recovery for input noise models. Here our necessary conditions say that if $m = o\left ( {n \log(n) \over \snr  } \right )$ then recovery would be
impossible.  Evidently, either the $\snr$ or the number of measurements must scale linearly with $n$ to ensure support recovery. Thus either we must operate in an essentially
noiseless regime or forsake all compression. We also extend our results to approximate support recovery. Here a tradeoff between the number of measurements, $\snr$ and
support errors for different sparsity ratios. These tradeoffs are summarized in Column 2 of Table~\ref{tab:SNR_dist_tradeoff}. An interesting aspect of these results is that a constant
$\snr$ is sufficient if we could tolerate a constant fraction of errors in the support recovery. To establish the necessary conditions we use Fano's inequality and its
variations \cite{Ibragimov81}. For deriving sufficient conditions we analyze the performance of the Maximum-Likelihood (ML) estimator based on a novel insight that every large
support error event is essentially contained in the union of single support error events. This leads to a sharp bound that is order-wise optimal. Our necessary and sufficient
conditions for different sparsity levels require similar scaling of $\snr$, and the number of measurements(see Table~\ref{tab:LASSOvsML}).

\textbf{Related Literature}- The necessary condition that $\snr = \Omega(\log(n))$ irrespective of the number of measurements was first reported by the authors in
\cite{AeronITA08}. This paper extends these results to include necessary conditions on the number of measurements. Similar conditions have also been reported by Fletcher et. al.
\cite{Fletcher09} but due to the constraints imposed on the signal space---the signal is limited to have small amplitude variations on its support elements---their conditions
are conservative (see discussion in \cite{SrvZhao_arxiv}) for our setup. Necessary conditions have also been derived by Wainwright~\cite{WainwrightITDec09}. When the bounds of  \cite{WainwrightITDec09} (see Theorem 2 in \cite{WainwrightITDec09}) are applied to our setup, it implies that the number of measurements scale as $\Omega(\log(n))$, which is conservative. In addition \cite{WainwrightITDec09}, primarily imposes conditions on the number of measurements but does not impose separate bounds on $\snr$. In contrast we show that unless $\snr$ scales as $\Omega(\log(n))$ support recovery is impossible regardless of $m$. Furthermore, for $\snr =
O(\log(n))$ we show that $m$ must scale as $\Omega(k \log(n/k))$. Sufficient conditions for support recovery for output noise models has been described in
\cite{WainwrightITDec09,Mehmet08,Fletcher09,Candes07,WainwrightITMay09} as well. Nevertheless, these upper bounds are also significantly weaker than that appearing here.
Both Wainwright \cite{WainwrightITDec09} and Akcakaya et. al. \cite{Mehmet08} use union bounding to derive error bounds for exact recovery. Union bounds are generally
conservative and results in requiring significantly high $\snr$, i.e. significantly low admissible noise variance (see for instance, Theorem 1 in \cite{WainwrightITDec09}). The sufficient
conditions of Fletcher et. al. \cite{Fletcher09} is based on Greedy Basis Pursuit algorithm. However, their analysis, as described earlier, constrains the signals, $\bX$, to
have small amplitude variations on its support elements and when applied to our output noise setup is conservative (see again discussion in \cite{SrvZhao_arxiv}). While
\cite{Reeves08,Mehmet08} derive some results for approximate support recovery, the achievable region in terms of number of measurements and $\snr$ as a function of achievable
distortion is implicitly stated and is therefore not comparable to the results presented here.

\begin{table*}
\begin{center}
\begin{tabular}[ht]{|c|c|c|}
     \hline
\multicolumn{3}{|c|}{EXACT SUPPORT RECOVERY (Output Noise Model)}\\
\hline \multirow{2}{*}
    & Linear Sparsity & Sub-Linear Sparsity   \\
&$0 < \alpha = \alpha_n=\frac{k}{n}$ &$\alpha_n=\frac{k}{n} = n^{-\gamma},\,\,\gamma>0$\\
  \hline
  \multirow{2}{*}
{Necessity (this paper)} & $\snr = \Omega\left({\log(n) \over \beta^2}\right)$ & $\snr = \Omega\left({\log(n) \over \beta^2}\right)$     \\
&$m =\Omega(n)$ & $m = \Omega(k \log\frac{n}{k})$\\
\hline
  \multirow{2}{*}
{Sufficiency (this paper)} & $\snr = {32\log(2n) \over \beta^2}$ & $\snr = {32\log(2n) \over \beta^2}$     \\
&$m =6n H_2(2\alpha)$, $\alpha \leq 0.04$ & $m = 6k \log\frac{n}{2k}$\\
\hline
\end{tabular}
 \caption{\small Summary of fundamental bounds for exact support recovery in the worst-case setting described in Equation~(\ref{eq:model_snr}). $\beta$ is the minimum absolute value of the signal $\bX$ on its
 support set; $H_2(\cdot)$ denotes the binary entropy function; $k$ is the maximum allowable cardinality (sparsity) of the support of $\bX$; $\alpha$ is the maximum
 sparsity ratio and; $1/\snr$ is the noise variance in each noise dimension. The necessary conditions are stated for arbitrary (not necessarily IID) matrices, $\bG$, such that the marginal distribution of each component has zero mean and variance $1/m$. The sufficient conditions are stated for the
 case when each element of $\bG$ is drawn IID $\sim {\cal N}(0,\frac{1}{m})$. %Note that the necessary conditions have to be interpreted as follows:
 %The minimum SNR for support recovery  must scale as $\snr = \Omega(\log(n))$ and for $\snr = \Theta(\log(n))$ the measurements must scale as
 %$m = \Omega(k \log(n/k))$
 } \label{tab:LASSOvsML}
\end{center}

\end{table*}

\subsection{Rate distortion bounds}
In the second part of the paper, we consider sparse Bayesian signal models for $\bX$ to fully exploit the power of information theoretic methods. This naturally leads us to
characterizing necessary and sufficient conditions in terms of the rate distortion function.

We first consider \emph{arbitrary pointwise distortion metrics}, i.e., $\frac{1}{n} d(\hat \bX, \bX) = \frac{1}{n} \sum_j d(\hat X_j,X_j),\,\,j = 1,\,2,\,\ldots,\,n$, where
$X_j,\,\hat X_j$ are the j-th components of $\bX,\,\hat \bX$ respectively. For deriving necessary conditions we develop a new modified Fano's inequality that provides us with a
\emph{worst case} lower bound to the probability of error in reconstruction to within a distortion $\frac{1}{n} d(\hat \bX, \bX) \leq d_0$ in terms of the scalar rate distortion
function $R_{X}(d_0)$ and mutual information $\ind(\bX,\bY)$, between $\bX$ and $\bY$. This bound is of independent interest since it can be applied to non-sparsifying
distributions as well. In particular we show that,
\begin{align*}
\prob \left( \frac{1}{n}d(\hat{\bX},\bX) \geq d_0\right) \geq \dfrac{R_{X}(d_0) - c_0 - \frac{1}{n}\ind(\bX;\bY)}{R_{X}(d_0) }
\end{align*}
for some small constant $c_0 < R_{X}(d_0)$.

For deriving sufficient conditions we compute upper bounds to the probability of error subject to a tolerable distortion based on the so called \emph{covering property} of rate
distortion theory. In particular we formalize a minimum distance decoder (distance measured in terms of given distortion metric) over the set of rate distortion quantization
points. We then specialize our bounds to the mean squared distortion metric. The results are summarized in the second column of Table~\ref{tab:SNR_dist_tradeoff}. Our necessary
and sufficient conditions for the number of measurements and $\snr$ match within a constant factor for the linear sparsity regime.

\textbf{Related Literature}- Rate distortion analysis has been reported in \cite{Fletcher06,Vivek_SSP07} for mean squared error and for a Gaussian source. In contrast our
expressions apply to general distortion measures and to any source for which a rate distortion function is defined. These results appeared in our preliminary
work~\cite{aeronITW07}. In addition the results in \cite{Fletcher06} for the case when $\bG$ is random are only proven for $k=1$. In contrast in this paper we prove results for
general $k = \alpha n$. For a fixed problem size $(n,k,m)$ the results in \cite{Vivek_SSP07} are stated in terms of a critical $\snr$ threshold. This makes the expressions
implicit in the number of measurements required as a function of signal sparsity and therefore the scaling laws are unclear.

\begin{table*}
\begin{center}
\begin{tabular}[ht]{|p{3.5in}|p{2.85in}|}
     \hline
 \multicolumn{2}{|c|}{APPROXIMATE SUPPORT RECOVERY - Sufficient conditions (Linear Sparsity Regime)}\\
 \hline
   Support Error Distortion $\frac{1}{k}\sum_{j=1}^n |I_{\{X_i\not=0\}}-I_{\{\hat X_i\not=0\}}|\leq d_0$ &  Mean Squared Distortion $\frac{1}{n} \|\hat \bX - \bX\|^2\leq d_0$ \\
  \hline
  $\snr
 =\Omega\left (\frac{H_2\left(2\alpha_n d_0\right)}{\beta^2}\right )$ and $m =\Omega( n H_2(2 \alpha_n))$ & $m = \Omega ( n R_{X}(d_0/2))$, for  $\snr = \Omega (\frac{R_{X}(d_0/2)}{d_0}) $.  \\
  \hline
\end{tabular}
\caption{\small The first column describes the achievable rate regions for approximate support recovery. Support error distortion $d_0$ is the fraction of the true support in
error. The second column describes the results for the Bayesian set-up in terms of the scalar rate distortion function for varying mean squared distortion. Here the distortion
$d_0$ is the desired mean-squared-distortion.} \label{tab:SNR_dist_tradeoff}
\end{center}
\end{table*}

The rest of the paper is organized as follows. In Section \ref{sec:prob_setup} we present our problem set-up. Here the notion of Sensing Capacity is introduced to study the
asymptotic behavior of both the output noise and input noise models. Section \ref{sec:scap_support} presents necessary and sufficient conditions for support recovery. In Section
\ref{sec:average_dist} we consider the Bayesian setup and derive bounds for signal recovery under \emph{arbitrary} distortion measures. This requires us to generalize the
traditional Fano's inequality to general (average) distortion measures and continuous signal spaces. We also provide extensions of Fano's inequality for discrete signal spaces
with Hamming distortion in reconstruction. Section \ref{sec:peupperbound} presents a novel ML upper bound for signal recovery to within a given squared distortion level. Using
these results, in Section \ref{sec:scap_bounds} we evaluate bounds for $\snr$ and number of measurements required to reconstruct $\bX$ to different levels of distortion level for output and input noise models. We then comment on the differences between worst-case and Bayesian setups.

\section{Problem Set-up}
\label{sec:prob_setup}
We consider output and input noise models described in Equations~(\ref{eq:model_snr}) and (\ref{eq:model_snr1}). The sparsity of $\bX$ is modeled both
deterministically and stochastically as is the compression matrix $\bG$. We use bold-face to denote vectors and matrices, while regular font is used to denote scalar components of the vector and
matrices. The jth component of a vector $\bX$ is denoted $X_j$, the jth column of a matrix $\bG$ is denoted $\bg_j$ and its ij-th component is denoted as $g_{ij}$. The cardinality of a set $S$ is denoted by $|S|$.
Given a set $S \subset \{1,\,2\,\ldots,\,n\}$, $\bX_S$ denotes the signal,
$\bX$, restricted to the set of components indexed by $S$. Similarly, we denote by $\bG_S$ the matrix formed from columns indexed by $S$. We use $\Pr(\cdot)$ and $\prob(\cdot)$ interchangeably to denote the probability of an event.
\paragraph{Non-Random Sparsity Signal Model:} We say that $\Xi^{\{k\}} \subset \Real^n$ is a family of k-sparse sequences if for every $\bX \in \Xi^{\{k\}}$,
the support of $\bX$ is smaller than or equal to $k$.
Formally, let
\begin{align*}
\mbox{\supp}(\bX) = \{j \mid X_j \not = 0\}
\end{align*}
Then $\Xi^{\{k\}}$ is a family of k-sparse sequences if,
\begin{align} \label{e.nonrand}
\Xi^{\{k\}} = \left\{ \bX: |\supp(\bX)| \leq k \right\}
\end{align}
We will refer to the ratio, $\alpha_n = k/n$ as the sparsity ratio. We will often work with subsets of $\Xi_{\beta}^{\{k\}} \subset \Xi^{\{k\}}$. These are sequences whose minimum absolute value is bounded away from zero by a constant $\beta \geq 0$:
\begin{align}
\label{e.nonrand1} \Xi_{\beta}^{\{k\}} = \{ \bX \in \Real^{n}: |\text{Supp}(\bX)| \leq k,\,\,|X_j|\geq \beta, \,\,\,\forall\, j \in \supp(\bX) \}
\end{align}

We will see when we derive necessary conditions that $\beta>0$ is necessary for support recovery. This is mainly because it is impossible to determine the support of a signal
with arbitrarily small components under noisy measurements.  This condition is also assumed by other authors \cite{Wainwright,Candes07}.

We denote by $\Xi^{k}\subset \Xi^{\{k\}}$ the set consisting of \emph{exactly} k-sparse sequences.
\begin{align} \label{e.nonrand_exact}
\Xi^{k} = \left\{\bX :  |\supp(\bX)| = k \right\}
\end{align}
This distinction is important and the reader should keep this in mind. The subset $\Xi_{\beta}^{k} \subset \Xi^{k}$ is analogously defined.

\paragraph{Bayesian signal model:} We say that a prior distribution on $\bX$ is an asymptotically sparsifying distribution if for sufficiently large $k,n$ the distribution concentrates
all the measure on a subset of $\Xi^{\{k\}}$. In this paper we will provide general results for arbitrary sparsifying priors and explicit bounds for the following Gaussian mixture model, namely, each component of the signal is distributed as:
\begin{align*}
X_i \stackrel{d}{\sim} P_X = \alpha {\cal N}(\mu_1,\sigma_{1}^2) + (1 - \alpha) {\cal N}(\mu_0,\sigma_{0}^{2})
\end{align*}
The corresponding $n$ dimensional distribution of $\bX$ is realized as a product measure on $\Real^n$. As an example note that for $\mu_1 = 1, \mu_0 = 0$ and  $\sigma_{1}=
\sigma_0 \rightarrow 0$ this mixture model asymptotically models binary sparse sequences with sparsity highly concentrated around $k = \alpha n$. The main reason for using a Bayesian
signal model is that it lends itself to information theoretic tools and allows us to study the tradeoffs between the number of measurements at different distortion levels for a given $\snr$.

\subsection{Sensing Capacity}

The nature of the results developed in the paper are asymptotic, namely, we let the signal dimension $n$ and the sparsity $k$ each approach infinity at different rates and
derive bounds on the number of measurements, $m$, and $\snr$, for exact/approximate reconstruction of $\bX$. In this context we also derive bounds for $m$ and $\snr$ for reconstruction of functions $\bZ=f(\bX)$ of $\bX$. For instance, we consider functions $f(\cdot)$ that indicate the support or sign function of $\bX$. We denote $\hat \bX(\bY)$ (resp. $\hat \bZ(\bY)$) as an estimate of $\bX$ (resp $\bZ$) based on the observation $\bY$. The
distortion between the estimate $\bZ$ and the estimate $\hat \bZ$ is denoted by $\frac{1}{n} d(\hat \bZ,\bZ) = \frac{1}{n} \sum_{j} d(\hat Z_j ,Z_j) $ for some scalar distortion
metric $d(\cdot,\cdot)$.

The sensing capacity involves determining the largest ratio $\frac{n H_2(\alpha_n)}{m} = \frac{n H_2({k \over n})}{m}$, required for reconstruction to within a desired
distortion. To build motivation on this ratio, consider again the maximum sparsity ratio $\alpha_n = \frac{k}{n}$.  The cardinality of the support set is $2^ {\log(\sum_{j=0}^k
{n \choose j})}) = O(2^{n H_2(k/n)})$, where $H_2(\cdot)$ denotes the binary entropy function \cite{Cover}. The term $n H_2(k/n)$ is a measure of the entropy of the support set,
i.e., the average number of bits required to uniquely encode the support set. The sensing capacity measures the number of source bits/measurement required for accurate decoding to a desired distortion level from compressed measurements.

If sensing capacity is a constant, it implies that the number of measurements required is proportional to the source entropy. On the other hand if the sensing capacity approaches zero, it means that the number of measurements must increase significantly faster than the source entropy. This also implies that the compression operator $\bG$ offers poor compression.

%For sub-linear sparsity $H_2(k/n) \rightarrow 0$, while for linear sparsity, $H_2(k/n)$ is bounded away from zero. In essence sub-linear  sparsity can be associated with a
%\emph{vanishing} entropy rate of the source support in the limit.

We next define the $\epsilon$-sensing capacity for a signal $\bX$ of dimension $n$ and  with maximum sparsity $k$. We use $\Xi$ to denote a suitable subset of admissible
signals, $\bX$. This could be any subset such as those described in Equations (\ref{e.nonrand1}) and (\ref{e.nonrand_exact}).
\begin{align} \label{e.nrscapeps}
C_{n,\epsilon}^{1}(\snr,\alpha_n,d_0)\stackrel{\Delta}{=}C_{n,\epsilon}^{1}(\snr,k,d_0) = \sup_m\left\{\frac{nH(k/n)}{m}:  \ex_\bG \sup_{\bX \in \Xi} \prob\left(\frac{1}{n} d(\bZ,{\hat \bZ}) \leq d_0| \bG, \bX\right) \geq 1 -
\epsilon \right\}
\end{align}
where the probability is over $\bN$. Note that one may choose a less conservative  notion by interchanging the order of $\max_{\bX\in \Xi^{\{k\}}}$ and $\ex_\bG$:
\begin{align} \label{e.nrscapeps1}
C_{n,\epsilon}^{2}(\snr,\alpha_n,d_0)\stackrel{\Delta}{=}C_{n,\epsilon}^{2}(\snr,k,d_0) =  \sup_m\left\{\frac{nH(k/n)}{m}:   \sup_{\bX \in \Xi} \ex_\bG \prob\left(\frac{1}{n}
d(\bZ,{\hat \bZ}) \leq d_0| \bG, \bX\right) \geq 1 - \epsilon \right\}
\end{align}
For the Bayesian set-up the sensing capacity is defined as,
\begin{align}
\label{e.nrscapeps2} C_{n,\epsilon}^{3}(\snr,\alpha_n,d_0)\stackrel{\Delta}{=}C_{n,\epsilon}^{3}(\snr,k,d_0) =  \sup_m\left\{\frac{nH(k/n)}{m}:  \ex_{\bG,\bX}
\prob\left(\frac{1}{n} d(\bZ,{\hat \bZ}) \leq d_0| \bG, \bX\right) \geq 1 - \epsilon  \right\}
\end{align}
where the probability is again over $\bN$. Since
\[ \ex_\bG \sup_{\bX \in \Xi^{\{k\}}} \prob\left(\frac{1}{n} d(\bZ,{\hat \bZ}) \geq d_0| \bG, \bX\right) \geq \sup_{\bX \in \Xi^{\{k\}}} \ex_\bG \prob\left(\frac{1}{n} d(\bZ,{\hat \bZ}) \geq d_0| \bG,
\bX\right) \geq \ex_{\bG,\bX} \prob\left(\frac{1}{n} d(\bZ,{\hat \bZ}) \geq d_0| \bG, \bX\right) \]
This implies that
\begin{equation} \label{e.relationscap}
C_{n,\epsilon}^{1}(\snr,k,d_0) \leq C_{n,\epsilon}^{2}(\snr,k,d_0) \leq C_{n,\epsilon}^{3}(\snr,k,d_0)
\end{equation}
This chain of inequalities implies that an upper bound for the Bayesian sensing capacity is an  upper bound for the other notions as well. A lower bound for the worst-case
sensing capacity (Equation~(\ref{e.nrscapeps})) is a lower bound for the other notions as well. To derive the lower bound to sensing capacity we derive an upper bound on the
probability of error using Maximum Likelihood (ML) analysis that uniformly holds for all $\bX \in \Xi^{\{k\}}$. For this reason we primarily focus on the notion of Equation
(\ref{e.nrscapeps}) and  Equation (\ref{e.nrscapeps2}). To avoid cumbersome notation we drop the superscript denoting the different notions, namely, we employ
$C_{n,\epsilon}(\cdot) \stackrel{\Delta}{=} C_{n,\epsilon}^i (\cdot)$, since it is usually clear from the context.

We propose an asymptotic definition for sensing capacity by letting $n\longrightarrow \infty$ as follows.

\begin{defn}
Let $\{\alpha_n\}$, be any sequence of sparsity ratios where $k$ is either fixed or approaching infinity linearly or sub-linearly with $n$.  \emph{Sensing capacity} is the
supremum over all the sensing rates such that as the signal dimension, $n$, the number of measurements, $m$, and the dimension of the (possibly) random sensing matrix, $\bG \in
\Real^{m\times n}$, approaches infinity, there exists a sequence of estimators $\hat \bZ$ such that the probability that the distortion, $\frac{1}{n} d(\bZ,\hat \bZ)$ is below
$d_0$ approaches one. Formally,
\begin{align*}
C(\snr,\{\alpha_n\},d_0) & = \lim_{\eps \rightarrow 0}\limsup_{m,n} C_{n,\epsilon}(\snr,\alpha_n,d_0)
\end{align*}
where we explicitly denote the dependence of capacity on $\snr$, sparsity sequence $\alpha_n$, and distortion level $d_0$.
\end{defn}

In the following we begin by considering the case of exact support recovery for the family of $k$-sparse sequences.

\section{Support Recovery: Worst-Case Setting}
\label{sec:scap_support}

In this section we consider the problem of exact support recovery under the models  of Equations~(\ref{eq:model_snr}) and  (\ref{eq:model_snr1}) for the non-random parameter
set, $\Xi_{\beta}^{\{k\}}$ given by Equation~(\ref{e.nonrand1}). Suppose, $\hat \bX$ is the estimate for $\bX$ based on measurements $\bY$. Recall that by exact support recovery we
mean that,
\begin{align*}
\prob_e = \ex_{\bG}  \sup_{\bX \in \Xi_{\beta}^{\{k\}}}  \prob\{\supp(\hat \bX) \not = \supp(\bX)\mid \bX,\bG \} \longrightarrow 0
\end{align*}
where the probability is over $\bN$. In this context one may also talk about sign pattern recovery,
\begin{align*}
\prob_e = \ex_{\bG} \sup_{\bX \in \Xi_{\beta}^{\{k\}}} \prob\{\sgn(\hat \bX) \not = \sgn(\bX) \mid \bX, \bG \} \longrightarrow 0
\end{align*}
 Here the $\sgn$ function is described by
\begin{align*}
\sgn(X) = \left \{ \begin{array}{c} 1,\,\, \mbox{if}\,\, X > 0 \\
-1,\,\, \mbox{if}\,\, X < 0 \\ 0,\,\, \mbox{if}\,\, X = 0
\end{array} \right .
\end{align*}
It is easy to see that the results derived below also hold for sign pattern recovery with appropriate adaptation of the proof methodology and the subsequent results only differ
by constant factors and in particular does not change the resulting \emph{scaling laws}. Therefore we will focus on the problem of support recovery. For this set-up following
are our main results for the output and input noise models.

\begin{thm}[Output Noise Model:Necessity]\label{thm:mainsuppthm1} Consider the output noise model of Equation~(\ref{eq:model_snr})
with the signal set defined by Equation~(\ref{e.nonrand1}). Let $\bG$ be any matrix such that the marginal distribution for each component has zero mean with variance $\frac{1}{m}$. Then there exists no estimator that can recover the support if $ \snr = o(\log(n))$. Furthermore, for $\snr=O(\log(n))$ support recovery is impossible if $m =o(k\log(n/k))$.
\end{thm}
The proof can found in Section 3.1.2. Note that we do not have to assume that the components of the sensing matrix are distributed IID. The proof of the theorem also shows that the number of measurements can not be decreased significantly unless $\snr$ scales as $n^{\gamma}$ for some $\gamma>0$. It is interesting to point out that in contrast to the noiseless case where $2k+1$ are required for signal reconstruction, the presence of even small noise (namely, variance scaling as $1/\log(n)$) significantly alters this fundamental bound.

The following result characterizes a partial converse of Theorem~\ref{thm:mainsuppthm1}.

\begin{figure}[t]
\begin{centering}
\includegraphics[width = 4 in]{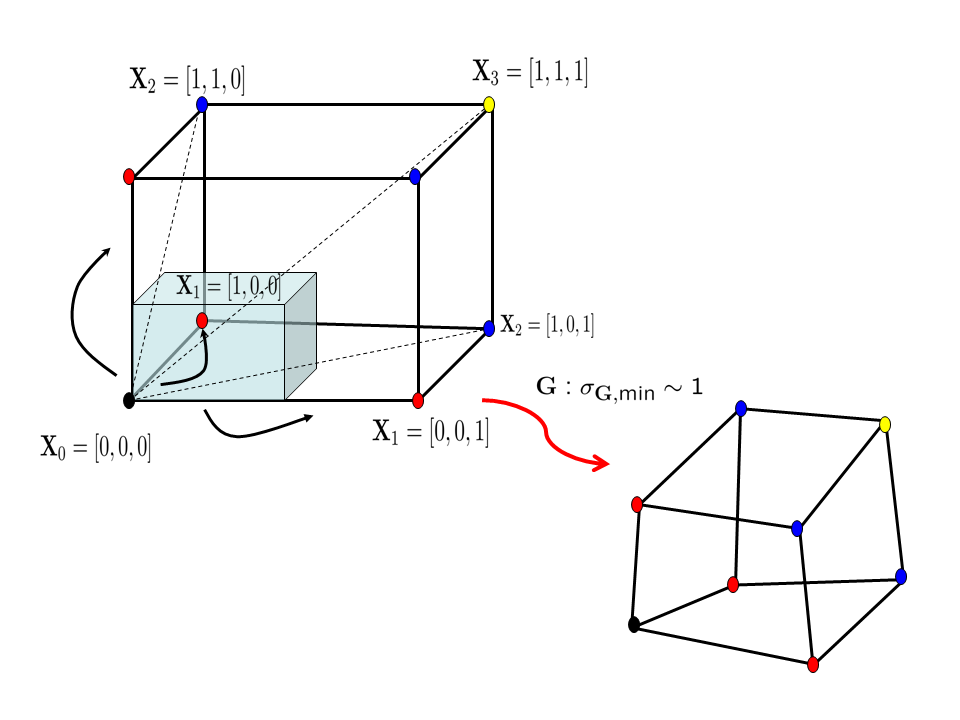}
\caption{\small Figure illustrating the intuition behind our ML analysis for support recovery using binary $\bX$ as an example. In the Figure $\bX_0$ is the true signal that is
taken to be the origin. Support error events with support errors more than $1$ are contained in union of events with support error of $1$ before the sensing/compression operator $\bG$ is applied. This property is essentially preserved under the transformation by $\bG$ if the minimum singular value of matrix $\bG$ is well behaved.  } \label{fig:tight_ML}
\end{centering}
\end{figure}

\begin{thm}[Output Noise Model:Sufficiency]\label{thm:mainsuppthmsuff1}
Suppose the sensing matrix, $\bG$, in Equation~(\ref{eq:model_snr}) is drawn from an IID Gaussian ensemble with each component  $g_{ij} \stackrel{d}{\sim} {\cal
N}(0,\frac{1}{m})$ and the signal set is given by Equation~(\ref{e.nonrand1}). If $m=\Omega(n H_2(\frac{k}{n}))=\Omega(k\log(n/k))$ and $SNR=\Omega(\log(n))$ then the ML
algorithm can exactly recover the support with high probability for all $\frac{k}{n} = \alpha_n \leq .04$. Alternatively, for any sensing matrix $\bG$ with $m \geq 2k + 1$ and
$SNR=\Omega(\log(n))$ the ML algorithm can recover the support with high probability, if the minimum singular value, $\sigma_{\bG,\min} = \min_{\bX \in \Xi^{\{2k\}} }
{\|\bG\bX\|_2^2 \over \|\bX\|_2^2}$ is bounded way from zero.
\end{thm}
\begin{remark}
Note that Theorem \ref{thm:mainsuppthmsuff1} for the deterministic case requires $\sigma_{\bG,\min}$ to be bounded away from zero. One may question whether this requirement is fundamental. We argue that
this is so here. Note that the optimal decoder must compare different signals with supports smaller than k and pick the most likely. If $\sigma_{\bG,\min}$ is arbitrarily small, it implies that
there are $k$ columns which are badly conditioned. In the presence of noise a worst-case signal emanating from these $k$ sparse columns will go virtually undetected relative to
noise.
\end{remark}
The proof for the deterministic and stochastic sensing matrices appear in Sections~\ref{sec:suffD} and \ref{sec:suffG}. A geometric intuition of the proof for deriving the sufficient condition is shown in Figure~(\ref{fig:tight_ML}) for binary $\bX$. The proof is based on the fact that for Gaussian noise $\bN$, before the compression operator $\bG$ is applied, the support errors larger than one are contained in the union of events with support error equal to one. We show that this is largely true when the compression is applied as well.
The proof for the random Gaussian matrix $\bG$ is based on the deterministic case.
 It turns out that the sparsity ratio $\alpha_n <0.04$ controls the singular values of the sub-matrix, namely, we can ensure $\sigma_{\bG,\min}>0$ with high probability for sparsity ratios below this number.

Note that we can also state these results in terms of sensing capacity. Formally, given any $\eps > 0$, there is an $n(\eps)$ such that for all $n \geq n(\eps)$ and any monotonic sequence
$\alpha_n<0.04$, there are positive constants, $c_1,\,c_2$, so
\begin{align*}
0< c_1 \leq C_{n,\eps}(\log(n),\alpha_n,0) \leq c_2
\end{align*}
In contrast to the optimistic results for output noise models, we have the following pessimistic result for the input noise model whose proof can be found in Section~\ref{sec:inpnec}.
\begin{thm}[Input Noise Model:Necessity]\label{thm:mainsuppthm2} Consider the input noise model of Equation~(\ref{eq:model_snr1}) with the signal set defined by Equation~(\ref{e.nonrand1}) and $\bG$
drawn from an IID Gaussian ensemble with each component $g_{ij}\stackrel{d}{\sim}{\cal N}(0,1/m)$. Let $\alpha_n$ be any positive monotonic sequence of sparsity ratios. Then recovery fails if $m =
o\left ( {n \max(\log(n),\log({1 \over \alpha_n})) \over \beta^2 \snr  } \right )$. Alternatively, the sensing capacity is zero.
\end{thm}
This says that for the input noise model one cannot expect meaningful compression in a noisy regime. To  ensure support recovery either the $\snr$ has to  scale linearly with
$n$, which implies essentially a noiseless regime, or the number of measurements must scale linearly with $n$ with any meaningful level of noise. This calls into question the
sensor network motivated compression schemes such as those presented in \cite{NowakIT} where the raw noisy measurements are randomly projected and transmitted to a fusion
center.
\subsubsection{Achievable Distortion Regions for Support Recovery}
In this section we will describe results for approximate support recovery, namely, we allow some distortion in support recovery. An important implication of our result is that in the constant sparsity regime it is sufficient for $\snr$ to be a constant independent of $n$ if we
accommodate a constant fraction of support errors. We account for the support distortion as $$d(\hat{ \bX},\bX)) = \frac{1}{k}\sum_{j=1}^n |I_{\{X_i\not=0\}}-I_{\{\hat X_i\not=0\}}|$$
where, $I_{\{\cdot\}}$ is the indicator function.

\begin{thm}
\label{thm:dist_supp} Consider the observation model of Equation~(\ref{eq:model_snr}) with $\bG$ drawn from a Gaussian ensemble.  Let $\bX \in \Xi_{\beta}^{\{k\}}$ and let $d_0$
be as described above. It follows that if $\snr \geq \frac{64 H_2\left(\frac{2k d_0}{n}\right)}{\beta ^2}$ and $m \geq 6 n H_2\left(2 \frac{k}{n}\right)$ the probability of
support error greater than distortion $d_0$ goes to zero. Consequently, it follows that for support recovery with constant distortion, $d_0$, in the linear sparsity regime, i.e,
$\alpha_n = k/n \geq \alpha>0$, it is sufficient for the $\snr$ to be a constant independent of the signal dimension $n$.
\end{thm}

\begin{proof}
The proof is based on the proof of Theorem~\ref{thm:mainsuppthmsuff1} and we refer the reader to the appendix.
\end{proof}

Note that Theorem \ref{thm:dist_supp} only trades off $\snr$ with the distortion. However one would expect that with allowable distortion in support recovery it is possible to
tradeoff number of measurements with distortion. In the following sections we will develop this tradeoff of number of measurements with the rate-distortion function by
considering a Bayesian set-up. The main reason why this tradeoff is possible in a Bayesian set-up is due to the fact that before we analyzed a \emph{worst case}
set-up while in Bayesian case we analyze an \emph{average case} scenario and it turns out that on an average the number of measurements can indeed be traded off with distortion.

\subsection{Proof of Theorems~\ref{thm:mainsuppthm2} and \ref{thm:mainsuppthm1}: Necessary Conditions}
\label{sec:necessary_cond_support}
%In this section we establish the proofs for Theorem~\ref{thm:mainsuppthm2} and Theorem~\ref{thm:mainsuppthm1}.
We derive necessary conditions based on lower bounds to probability of error. As we pointed out in Equation~(\ref{e.relationscap}) putting a suitable measure on the signal $\bX$
can provide necessary conditions for the worst-case setup. This motivates employing different versions of Fano's Lemma to establish the results. The standard version of the
lemma appears in \cite{Cover} and we repeat it here for the sake of completion:
\begin{lem}
\label{lem:Fano_Cover} Suppose ${\cal X}$ is a finite discrete set and $\bX \in {\cal X}$ is distributed uniformly over this finite set. Let the observation $\bY$ be distributed according to the conditional distribution $\prob(\bY|\bX)$, with $\bX \in {\cal X}$.  let
$\hat{\bX}(\bY)$ denote the estimate of $\bX$ given $\bY$. Then  the probability of error in estimating $\bX$ from $\bY$  is lower bounded by,
\begin{align*}
\prob (\hat{\bX}(\bY) \neq \bX)\geq 1 - \frac{\ind(\bX;\bY) + \log 2}{\log (|{\cal X}|-1)}
\end{align*}
where $\ind(\bX;\bY)$ denotes the mutual information between $\bX$ and $\bY$.
\end{lem}

An alternate version of Fano's lemma stated in \cite{Yannis88} provides a lower bound for $N$-ary hypothesis testing.
\begin{lem}
\label{lem:yannis_fano} Let $({\cal Y},{\cal B})$ be a $\sigma-$field and let $\prob_1,\ldots ,\prob_N$ be probability measures on ${\cal B}$ thought of as induced by $N$ hypotheses
$\{1,\,2,\,\ldots,\,N\}$. Denote by $\theta(y)$ the estimator of the measures defined on ${\cal Y}$. Then
\begin{align*}
\max_{1 \leq i \leq n} \prob_i(\theta(y) \neq \prob_i) \geq \frac{1}{N} \sum_{i=1}^{N} \prob_i(\theta(y) \neq \prob_i) \geq 1 - \frac{\frac{1}{N^2} \sum_{i,j} D(\prob_i\|
\prob_j) + \log 2}{\log (N-1)}
\end{align*}
where $\prob_i$ means the distribution conditioned on the hypothesis $i$ and $D(\prob_i \| \prob_j)$ is the  Kullback-Liebler (KL) distance between the distributions $\prob_i$
and $\prob_j$.
\end{lem}
Note that the use of these Lemmas requires a finite number of hypothesis or discrete alphabets.  Therefore, in order to use these Lemmas for general $k$-sparse sequences $\bX
\in \Xi_{\beta}^{\{k\}}$ we first show that the worst case probability  of error in support recovery is lower bounded by the probability of error in support recovery for $\bX$
belonging to $k$-sparse sequences in $\left\{0,\beta\right\}^n  $. To this end we have the following Lemma.
\begin{lem}
\label{lem:reduct} Let $\Xi_{\beta}^{\{k\}}$ be the family of $k$ sparse non-random sequences as defined  in Equation~(\ref{e.nonrand1}). Denote the conditional distribution of
$\bY$ given $\bX$ as $\prob(\bY \mid \bX)$. Let $\Xi_{\{0,\beta\}}^{\{k\}} = \{\bX \in \Xi_{\beta}^{\{k\}} \mid X_j = \beta,\,\, j \in \supp(\bX)\}$ be a subset of
$\Xi_{\beta}^{\{k\}}$ consisting of binary valued sequences. Let $\hat \bX$ denote an estimator for $\bX$ based on observation $\bY$. Then,
\begin{align} \nonumber
\prob_{e|\bG} = \min_{\hat \bX \in \Xi_{\beta}^{\{k\}}} \max_{\bX \in \Xi_{\beta}^{\{k\}}} \prob\{\supp(\hat \bX) \neq \supp(\bX) | \bG, \bX \}  &\geq \min_{\hat \bX \in
\Xi_{\beta}^{\{k\}}}\max_{\bX \in \Xi_{\{0,\beta\}}^{\{k\}}} \prob (\hat \bX \neq
\bX ,\,\,\hat \bX \in \Xi_{\{0,\beta\}}^{\{k\}} | \bG, \bX)\\
\geq & \min_{\hat \bX \in \Xi_{\{0,\beta\}}^{\{k\}}}\max_{\bX \in \Xi_{\{0,\beta\}}^{\{k\}}} \prob (\hat \bX \neq \bX ,\,\,\hat \bX \in \Xi_{\{0,\beta\}}^{\{k\}}| \bG, \bX) \label{eq:condprobnote}
\end{align}
\end{lem}
\begin{proof}
See Appendix.
\end{proof}
The main idea behind the proofs of the results that follow below is to first lower bound the error probability by using Lemma \ref{lem:reduct} and restrict attention to binary
sequences. Next we further restrict the signal class to a smaller subset of $\Xi_{\{0,\beta\}}^{\{k\}}$ of cardinality $n$. Then, finally using Lemma \ref{lem:yannis_fano} we
derive the lower bounds for the set of binary sequences. The lower bound thus obtained yields the necessary conditions.
\subsubsection{Input Noise Model(Proof of Theorem~\ref{thm:mainsuppthm2})} \label{sec:inpnec}
From Lemma \ref{lem:reduct} it is sufficient to focus on the case when $\bX$ belongs to the set of $k$-sparse sequences in $\left\{0,\beta\right\}^n$ and any subset of these
sequences. We will establish the first part of the Theorem as follows:- Let $\Xi_{\{0,\beta\}}^{\{\eta\}}$ be the subset of $\eta < k$ sparse binary valued sequences. Let $\bX_0
\in \Xi_{\{0,\beta\}}^{\{\eta\}}$, be an arbitrary element with support $\supp(\bX_0) = \eta-1$. Next choose $n$ elements $\bX_j, \, j=1,\,2,\,\ldots,\,n$ with support equal to
$\eta$ and at a unit Hamming distance from $\bX_0$. Denote by the probability kernel $\prob_j,\,0\leq j \leq n$ the induced observed distributions. Under the AWGN noise model,
for a given $\bG$, and a fixed set of elements, $\bX_j$, the probability kernels are Gaussian distributed, i.e.,
\begin{align*}
{\cal H}_j: \bY \stackrel{d}{\sim} \prob_j \equiv {\cal N} \left ( \bG\bX_j,\frac{\bsg}{{\snr}}\right ),\,\,j=0,\,1,\,\ldots,\,n
\end{align*}
where $\bsg = \bG\bG^T$. Furthermore we have $n+1$ hypotheses. Consider now the support recovery problem. It is clear that the error probability can be mapped into a
corresponding hypothesis testing problem. For this we consider $\theta(\bY)$ as estimate of one of the $n+1$ distributions above and we have the following set of inequalities.
\begin{align*}
\prob_{e\mid \bG} = \max_{\bX \in \Gamma^{\eta}} \prob_{\bX} ( \hat \bX \neq \bX \mid \bG) = \max_j \prob_j ( \theta(\bY) \neq \prob_j \mid \bG) \geq \frac{1}{n+1} \sum_{j=0}^n
\prob_j ( \theta(\bY) \neq \prob_j \mid \bG)
\end{align*}
where we write $\prob_{e \mid \bG}$ to point out that the probability of error is conditioned on $\bG$. Applying Lemma \ref{lem:yannis_fano} it follows that the probability of
error in exact support recovery is lower bounded by,
\begin{align*}
\prob_{e\mid \bG} \geq \frac{\log(n ) - \frac{1}{(n+1)^2}\sum_{i,j, i \neq j} D(\prob_i\| \prob_j) - \log 2}{\log (n)}
\end{align*}

We observe that under AWGN noise $\bN$ that,
\begin{align} \label{e.step1}
D(\prob_i\| \prob_j)  =  \snr (\bX_i - \bX_j)^T \bG^T \bsg^{-1} \bG(\bX_i - \bX_j) = \snr (\bX_i - \bX_j)^T \bV \left [ \begin{array}{cc} I & 0 \\0 & 0 \end{array}\right ] \bV^*(\bX_i - \bX_j)
\end{align}
where $\bsg = \bG\bG^T$, $\bG=\bU[\bLambda,\,\, 0] \bV^*$ is the SVD of $\bG$ with $\bV=[\bv_1,\bv_2,\,\bv_3,\,\ldots,\,\bv_n]=[v_{rs}]$. The last equality in
Equation~(\ref{e.step1}) follows from straightforward algebraic manipulations. Now by noting that $(\bX_i - \bX_j)$ is at most a 2-sparse vector with its non-zero entries equal
to $\beta$ at some locations $q$ and $p$, we can further reduce the last expression to $ D(\prob_i\| \prob_j)= \snr \beta^2 \sum_{l=1}^m (v_{pl}-v_{ql})^2$. Now using the
standard rotational invariance properties of IID Gaussian matrices ~\cite{Verdu_Book04}, that its singular vectors are uniformly distributed over a sphere, it follows by taking
expectations and using symmetry that,
\begin{align}
\label{e.step2} \prob_e = \ex_{\bG} \prob_{e \mid \bG} \geq \frac{\log(n ) - \frac{n}{n+1}\frac{ 2 \beta^2 \snr \, m }{n} - \log 2}{\log (n)}
\end{align}
Now, the error probability is bounded away from zero by $\epsilon$ if the number of measurements scales as follows:
$$
m = o \left ( {(n+1) \log(n) \over \beta^2 \snr} \right )
$$
To establish the second upper bound we consider the family,  $\Xi_{\{0,\beta\}}^{k}$ of exact k-sparse binary valued sequences which form a subset of
$\Xi_{\{0,\beta\}}^{\{k\}}$. Following similar logic as in the proof of the first part, for the set of exactly $k$-sparse sequences, we form the corresponding $\binom{n}{k}$
hypotheses. Then,
\begin{align}
\label{e.step3} \prob_e =  \ex_{\bG} \prob_{e \mid \bG} \geq \frac{\log (\binom{n}{k} - 1)- \frac{1}{\binom{n}{k}^2} \sum_{i,j, i \neq j} D(\prob_i\| \prob_j) - \log 2}{\log
(\binom{n}{k}-1)}
\end{align}
We compute the average pairwise KL distance,
\begin{align*}
& \frac{1}{{n \choose k}^2}\sum_{i,j, i \neq j} D(\prob_i\| \prob_j) \nonumber \\
& = \frac{1}{{n \choose k}}\sum_{j=1}^{k} \snr (\bX - \bX')^{T}\bG^T \bsg^{-1}\bG (\bX - \bX'). \sharp(\mbox{sequences  $\bX'$ at hamming distance 2j from $\bX$})
\end{align*}
The equality above follows from symmetry. Again using the standard rotational invariance properties of IID Gaussian matrices~\cite{Verdu_Book04}, the above equation implies that
,
\begin{align*}
\frac{1}{{n \choose k}^2}\sum_{i,j, i \neq j} D(\prob_i\| \prob_j) = \frac{m}{n} \frac{1}{{n \choose k}}\sum_{j=1}^{k} \snr \beta^2{ n -k \choose j} { k \choose j} (2j) =
\frac{m}{n} 2 \beta^2 \snr  \alpha_n n (1 - \alpha_n)
\end{align*}
where the last equality follows from standard combinatorial  identity. The proof then follows by noting that for large enough value of $n$, $\log (\binom{n}{k}- 1 ) \geq \alpha_n n
\log \frac{1}{\alpha_n}$.
\subsubsection{Output Noise Model (Proof of Theorem~\ref{thm:mainsuppthm1})}
We will now establish Theorem~\ref{thm:mainsuppthm1} namely, that if $SNR = o(\log(n))$ support recovery is impossible. Furthermore, if $SNR = O(\log(n))$ support recovery will
be impossible if the number of measurements scales as $o(k\log(n/k))$. The first part follows from the following Proposition.
\begin{prop}[Output noise model - SNR Bound]
\label{prop:cs_scap_support_ub} For the observation model of Equation~(\ref{eq:model_snr}) with the signal set of Equation~(\ref{e.nonrand1}) the $\snr$ must scale with
$\frac{\log(n)}{2 \beta^2}$ for perfect support recovery irrespective of which sensing matrix is used.
\end{prop}
\begin{proof}
The proof follows along the same lines as the proof  of Theorem \ref{thm:mainsuppthm2} with $\bsg = \mathbf{I}$ up  to Equation~(\ref{e.step1}). In the Kullback Leibler distance
calculation we are now left with the term $\bG^T\bG$. Since $\bG$ is normalized its expected value is identity. Therefore, we no longer get the factor $n/m$ in
Equation~\ref{e.step2}. Consequently, following the rest of the steps we have that, $2\beta^2 \snr \geq \log(n)$ for exact support recovery.
\end{proof}
Next we establish what happens for $SNR = O(\log(n))$ to prove the second part of Theorem~\ref{thm:mainsuppthm1}. First, note that  if the sparsity, $k$, grows linearly with the
signal dimension, $n$, there is nothing to prove, since it is well-known~\cite{Donoho1} that the number of measurements must scale at least as  $2k+1 = \Omega(n)$ even when there is no noise to guarantee
support recovery. For this reason we focus on the sub-linear case namely, $k = n^{-\gamma},\,\,\gamma<1$. We consider the subset $\Xi_{\{0,\beta\}}^{k}$ consisting of strictly
$k$-sparse sequences taking values in $\{0,\beta\}^{n}$. From Lemma \ref{lem:reduct} we see that it is sufficient to focus on this set. Applying Lemma \ref{lem:Fano_Cover} with
a uniform prior on the support set we get
\begin{align} \label{eq:Fano_supp_cover}
\max_{\bX \in {\cal X}} \prob(\hat{\bX} \neq \bX |\bX, \bG) \geq \prob(\hat{\bX} \neq \bX |\bG)  \geq 1 - \frac{\ind(\bX;\bY|\bG) + \log 2}{\log( |{\cal X}| -1)}
\end{align}
where ${\cal X} = \Xi_{\{0,\beta\}}^{k} \subset \{0,\beta\}^n$  is the discrete alphabet in which values of $\bX$ are realized. The first inequality follows because the
worst-case probability of error is larger than the Bayesian error.

Note that strictly speaking since we are interested in the support errors,  the probability of error events and the mutual information term must contain the support of $\bX$ as
the variable but since we are restricting ourselves to binary valued sequences $\bX \in \Xi_{\{0,\beta\}}^{k}$, knowing the support implies that we know $\bX$.

Now  $\log |{\cal X}| = \log \binom{n}{k}$ since there are $\binom{n}{k}$ such hypothesis consisting of all the possible support locations with cardinality $k$.   We will now
upper bound the mutual information term. It follows that,
\begin{align*}
\ind(\bX;\bY|\bG) &= h(\bY|\bG) - h(\bY|\bX,\bG) \leq h(\bY) - h(\bN) \stackrel{(a)}{\leq} \sum_{i=1}^{m} h(Y_i) - \frac{m}{2} \log \left (2\pi e \frac{1}{SNR} \right ) \\ &
\stackrel{(b)}{\leq} \frac{m}{2} \log\left (2\pi e \left (\frac{k \beta^2}{m} + \frac{1}{SNR}\right )\right ) - \frac{m}{2} \log(2\pi e \frac{1}{SNR}) = \frac{m}{2} \log(1 + \frac{k \beta^2 \snr}{m})
\end{align*}
where $h(\cdot)$ is the differential entropy; (a) follows from  the fact that the noise is Gaussian and the chain rule together with the fact that conditioning reduces entropy;
(b) follows from the fact that Gaussian distributions maximizes differential entropy. Now from Equation (\ref{eq:Fano_supp_cover}) it follows that the number of measurements
must satisfy,
\begin{align} \label{e.lbdmeassupp}
m \geq \dfrac{\log\left( \binom{n}{k} - 1\right)}{\log ( 1 + \frac{k \beta^2 \snr}{m}) + \frac{\log 2}{m}}
\end{align}
Next unless $\snr = \Omega(\log(n))$ we know from Proposition~\ref{prop:cs_scap_support_ub} that support recovery is  impossible. Hence we set $\snr=\log(n)$, which is the
minimum possible. We next establish the theorem by contradiction. To this end let the number of measurements scale as $m = \rho_n\log (\binom{n}{k})$, with $\rho_n \rightarrow
0$, then, by rearranging the terms in Equation~(\ref{e.lbdmeassupp}) we get
\begin{align} \label{eq:lbnd}
\log \left ( 1 + \frac{k \beta^2 \snr}{\rho_n\log (\binom{n}{k})}\right ) + \frac{\log 2}{\rho_n\log (\binom{n}{k})}  \geq \dfrac{\log\left( \binom{n}{k} - 1\right)}{\rho_n\log
(\binom{n}{k})}
\end{align}
Next note that the expression on the left can be simplified by noting  that $$\frac{k \beta^2 \snr}{\rho_n\log (\binom{n}{k})} = \Theta \left ( {1 \over \rho_n(1- {\log(k) \over
\log(n)})} \right )$$ while the expression on the right has the scaling $\Theta(\rho_n^{-1})$. Consequently, if maximum admissible sparsity,  $k$, grows sub-linearly with $n$
then $\log(1+\frac{k \beta^2 \snr}{\rho_n\log (\binom{n}{k})}) = \Theta(\log(1+ \rho_n^{-1}))$ and Equation~(\ref{eq:lbnd}) can never be satisfied since $\rho_n \rightarrow 0$. This shows that for
sub-linear cases recovery is impossible if $m = o(\log (\binom{n}{k}))= o(nH_2(\alpha_n))$.
\begin{remark} Note that unless $\snr$ scales as $n^{\delta}$ for some $\delta>0$ we will still need the measurements to scale as $\Omega(k\log(n/k))$ to guarantee support recovery. \end{remark}
\subsection{Proof of Theorem~\ref{thm:mainsuppthmsuff1}: Deterministic Case} \label{sec:suffD}
In this section we derive sufficient conditions for support recovery for the output noise model for any given \emph{arbitrary deterministic} matrix $\bG$ and for general noise
covariance $\bsg$. For the output noise model of Equation (\ref{eq:model_snr}), we assume that each column of the deterministic $\bG$ is normalized. Subsequently we specialize
these results to the case when $\bG$ is chosen from the Gaussian ensemble and with $\bsg = \mathbf{I}$.

To simplify the exposition we introduce several new variables. We associate each admissible signal, $\bX \in \Xi^{\{k\}}$ by its support, $S$. We denote by $\bX_S$ the signal,
$\bX$, restricted to the set of components indexed by $S$. Similarly, we denote by $\bG_S$ the matrix formed from columns indexed by $S$. Since the maximum sparsity level is $k$
the number of different support sets is equal to  $\sum_{j=1}^k {n\choose j} - 1$.  We index the different support sets as $S_{\omega}$ with $\omega \in {\cal I}=\left \{
0,\,1,\,2,\,\ldots, \sum_{j=1}^k {n\choose j} - 1\right \}$. Also we denote by $\bX_{S_{\om}}^{\min}$ the minimum absolute value of the components of the signal $\bX$ on the
support set $S_{\om}$, i.e., $\bX_{S_{\om}}^{\min} = \min \{|X_j| : j \in S_{\omega}\}$. Without loss of generality we assume that the true signal is $\bX_0$, the support set of
the true signal to be $S_0$ corresponding to $\omega=0$. We denote by $X_{0,j}$ the jth component of the true signal.

For any $\om \not = 0$, we denote the
overlapping support by, $S_{0,\omega}$, false detection by, $S_{0^c,\omega}$ and missed detection by, $S_{0,\omega^c}$, namely,
\begin{align*}
\mbox{Overlap} - &  S_{0,\omega} = S_0 \cap S_{\omega} \\
\mbox{False Alarms} - &  S_{0^c,\om} =S_{0}^c \cap S_{{\om}}\\
\mbox{Misses} - & S_{0,\omega^c} = S_0 \cap S_{\om}^c
\end{align*}
For a given noise covariance $\bsg$ the ML estimator is given by,
\begin{align*}
\hat{\bX} = \min_{\bX \in \Xi_{\beta}^{\{k\}}} (\bY - \bG \bX)^T\bsg^{-1}(\bY - \bG \bX)
\end{align*}
The above ML estimator is hard to analyze. In order to simplify the analysis we will consider a sub-optimal ML estimator. To this end consider the set, $\Xi_{\beta/2}^{\{k\}}$.
Clearly, $\Xi_{\beta}^{\{k\}} \subset \Xi_{\beta/2}^{\{k\}}$. We propose the following sub-optimal ML estimator,
\begin{align} \label{eq:relaxML}
\hat{\bX}=\arg\min_{\bX\in \Xi_{\beta/2}^{\{k\}}}\|\bY-\bG\bX \|^2
\end{align}
and report $\supp(\hat{\bX})$ as the final solution. Note that this estimator is sub-optimal since it is prone to more errors.  To see this note that we consider a larger signal
set and we ignore possible noise correlation $\bsg$ in our estimator. Consequently, the error probability in detecting the correct support can only be larger than the optimal ML
estimator. The performance of the relaxed estimator provides an upper bound for the performance of ML estimator. Hence, we can write,
\begin{align} \label{eq:relaxederror}
\prob^{ML}_{e \mid \bG} \leq \prob_{e\mid \bG} = \prob\left(\bN  : \min_{ \om \not = 0, \, \bX_{S_{\om}}^{\min} \geq \beta/2}\|\bY-\bG_{S_{\om}} \bX_{S_{\om}}\|^2\leq \min_{
\bX_{S_0}^{\min} \geq \beta/2}\|\bY-\bG_{S_0} \bX_{S_0} \|^2 \right)
\end{align}
Note that in the above expression $\bX_{S_0}$ is not the true signal, $\bX_0$, but any other signal whose support is identical to that of the true signal. We then have the
following result.
\begin{lem}
\begin{align*} \prob^{ML}_{e \mid \bG} \leq \prob_{e\mid \bG} \leq \prob({\cal E}_1)+ \prob({\cal E}_2) \end{align*}
where
\begin{align*}
{\cal E}_1=\{\bN  : \min_{ \om \not=0, \, \bX_{S_{\om}}^{\min} \geq \beta/2}\|\bY-\bG_{S_{\om}} \bX_{S_{\om}}\|^2\leq \min_{ \tilde{\bX}}\|\bY-\bG_{S_0} \tilde{\bX}
\|^2\}
\end{align*}
\begin{align*}
{\cal E}_2 = \left\{\bN: \|(\bG_{S_0}^T\bG_{S_0})^{-1}\bG_{S_0}^{T} \bN\|_\infty\geq \beta/2\right\}
\end{align*}
\end{lem}
\begin{proof}
First note the following qualitative points. In the event ${\cal E}_1$ we have replaced the constrained minimization on the R.H.S. of the inequality in the error event with an
unconstrained one. This will simplify the subsequent analysis as closed form expressions can be obtained. The event ${\cal E}_2$ captures the
probability that the unconstrained minimization in ${\cal E}_1$ is very far from the constrained one. Here we use the fact that the minimum component on the support of the true signal $\bX_0$ is greater than $\beta$. We also relax our ML estimator so that we find a best fit with any signal sharing the same support set, $S_0$, as $\bX_0$ but with $\bX_{S_0}^{\min} \geq \beta/2$. Now, denote
\begin{align*}
A \overset{\Delta}{=} \left\{ \bN  : \min_{ \om \not = 0, \, \bX_{S_{\om}}^{\min} \geq \beta/2}\|\bY-\bG_{S_{\om}} \bX_{S_{\om}}\|^2\leq \min_{ \bX_{S_0}^{\min} \geq
\beta/2}\|\bY-\bG_{S_0} \bX_{S_0} \|^2  \right\}
\end{align*}
\begin{align*}
B \overset{\Delta}{=} \left \{ \bN: \min_{ \bX_{S_0}^{\min} \geq \beta/2}\|\bY-\bG_{S_0} \bX_{S_0} \|^2=\min_{\tilde{\bX}}\|\bY-\bG_{S_0}\tilde{\bX}\|^2 \right \}
\end{align*}
Then we have
\begin{eqnarray*}
\prob_e = \prob(A)=\prob(A\cap B)+\prob(A\cap \bar{B})\leq \prob(A\cap B)+\prob(\bar{B})
\end{eqnarray*}
The Lemma then follows by noting that,
\begin{eqnarray*}
A\cap B&=&\left \{\bN: \min_{\bX_{S_{\om}}^{\min} \geq \beta/2, \om \neq 0}\|\bY-\bG_{S_{\om}}\bX_{S_{\om}}\|^2\leq \min_{ \bX_{S_0}^{\min} \geq
\beta/2}\|\bY-\bG_{S_0} \bX_{S_0} \|^2 \right \}\cap B\\
&=&\left \{\bN: \min_{\bX_{S_{\om}}^{\min} \geq \beta/2, \om \neq 0}\|\bY-\bG_{S_{\om}}\bX_{S_{\om}}\|^2 \leq \min_{\tilde{\bX}}\|\bY-\bG_{S_0}\tilde{\bX}\|^2 \right \} = {\cal
E}_1
\end{eqnarray*}
and,
\begin{align*}
\bar{B} = \left \{\bN: \min_{\bX_{S_0}^{\min} \geq \beta/2}\|\bY-\bG_{S_0}\bX_{S_0}\|^2\not=\min_{\tilde{\bX}}\|\bY-\bG_{S_0}\tilde{\bX}\|^2 \right \}\subset {\cal E}_2
\end{align*}
\end{proof}
From the above Lemma, it is sufficient to focus on events ${\cal E}_1$ and ${\cal E}_2$ separately. The following lemma provides a result that considerably simplifies the error
event ${\cal E}_1$. It turns out that the event ${\cal E}_1$ is a subset of the union of atomic events, namely,
\begin{lem} \label{lem:superp1}
For $m \geq 2k+1$,
\begin{align*}{\cal E}_1 \subseteq \tilde {\cal E}_1 =  \bigcup_{X \in \{ \beta/2, -\beta/2\}} \bigcup_{j=1}^n
\left\{\bN: 2 \bN^T \bg_j X \geq \sigma_{\bG,\min} |X|^2\right\}
\end{align*}
where, $\bg_j$ is the $j-th$ column of the matrix $\bG$ and
\begin{align} \label{eq:rip}
\sigma_{\bG,\min} = \min_{|S| \leq 2k} \sigma_{\min}(\bG_{S}^{T}\bG_{S})
\end{align}
where $\sigma_{\min}(\bG_{S}^{T}\bG_{S})$ denotes the minimum singular values of $\bG_{S}^{T}\bG_{S}$.
\end{lem}
\begin{proof}
See Appendix.
\end{proof}
We now have the following Lemma.
\begin{lem} \label{lem:E1}
Consider the output noise model for a deterministic matrix $\bG$ with $m \geq 2k+1$ and $\bN$ distributed as ${\cal N}(0,\bsg)$. The
probability of the error event ${\cal E}_1$ is upper bounded by,
\begin{align} \label{eq:bndE1}
\prob({\cal E}_1) \leq \exp \left\{- \sigma_{\bG,\min}^{2}\frac{ \lambda_{min}(\bsg^{-1})\beta^2 \snr}{ 32}\right\} \exp\{\log 2n\}
\end{align}
where $\lambda_{\min}(\bsg^{-1})$ is the minimum eigenvalue value of the matrix $\bsg^{-1}$.
\end{lem}
\begin{proof} See Appendix.
\end{proof}
We now have the following Lemma for the error event ${\cal E}_2$. Again note that the result applies to any matrix $\bG$ (not necessarily Gaussian).
\begin{lem} \label{lem:E2}
For the setup of Lemma~\ref{lem:E1}, we have,
\begin{align*}
\prob ({\cal E}_2)  \leq \exp \left\{-\sigma_{\bG,\min}^{2}\frac{ \lambda_{\min}(\bsg^{-1})\beta^2\snr}{8} + \log 2 n\right\}
\end{align*}
\end{lem}
\begin{proof} See Appendix.
\end{proof}
By combining Lemmas~\ref{lem:E1} and \ref{lem:E2} we can prove the deterministic case of Theorem~\ref{thm:mainsuppthmsuff1}. We state it as a proposition since we will refer to it later.
\begin{prop} \label{prp:Pe_general}
Consider the setup of Lemma~\ref{lem:E1}. Then for exact support recovery it is sufficient that $m \geq 2k+1$ and $\snr =
\Omega\left(\dfrac{1}{\sigma_{\bG,\min}^{2}}\dfrac{64
 \log 2 n }{\beta^2 \lambda_{\min}(\bsg^{-1}) }\right)$.
\end{prop}
\begin{proof}
From Lemmas \ref{lem:E1} and \ref{lem:E2} it follows that for $m\geq 2k+1$,
\begin{align*}
 \prob_{e|\bG} &\leq \exp\left\{- \sigma_{\bG,\min}^2\frac{ \lambda_{\min}(\bsg^{-1}) \beta^2 \snr}{ 32} + \log 2n\right\} +
\exp\left\{-\sigma_{\bG,\min}^{2} \frac{\lambda_{\min}(\bsg^{-1}) \beta^2\snr}{8} + \log 2n \right\} \\
& \leq 2 \exp\left\{- \sigma_{\bG,\min}^{2} \frac{ \lambda_{\min}(\bsg^{-1})\beta^2 \snr}{ 32}+  \log 2n\right\}
\end{align*}
Therefore for $\snr = 2\cdot \sigma_{\bG,\min}^{-2} \dfrac{32  \log 2 n }{\beta^2 \lambda_{\min}(\bsg^{-1}) }$ the probability of error $\prob_{e|\bG} \leq 2 e^{-\log 2 n }$.
Thus with $n\rightarrow \infty$, $\prob_{e|\bG}$ goes to zero as $\frac{1}{n}$. This implies that $\snr$ scaling of $\Omega\left(\sigma_{\bG,\min}^{-2}\dfrac{64 \log 2 n
}{\beta^2 \lambda_{\min}(\bsg^{-1}) }\right)$ is sufficient.
\end{proof}
%
%\begin{remark}
%Note that Proposition \ref{prp:Pe_general} requires $\sigma_{\bG,\min}$ to be bounded away from zero. One may question whether this requirement is fundamental. We argue that
%this is so here. Note that the optimal decoder must compare different $k$ sparse signals and pick the most likely. If $\sigma_{\bG,\min}$ is arbitrarily small, it implies that
%there are $k$ columns which are badly conditioned. In the presence of noise a worst-case signal emanating from these $k$ sparse columns will go virtually undetected relative to
%noise.
%\end{remark}
%
\subsection{Proof of Theorem~\ref{thm:mainsuppthmsuff1}: Gaussian Case} \label{sec:suffG}
We will now focus on sensing matrices, $\bG$, drawn from an IID Gaussian ensemble.  As in the deterministic case we need to bound the probabilities of events, ${\cal E}_1$ and
${\cal E}_2$. We will first focus our attention on event ${\cal E}_1$.

We point out that the proof for the deterministic case cannot be directly applied. First, note that $\sigma_{\bG,\min}$ of Equation~(\ref{eq:rip}) is now a random variable.
Therefore, we need to average over this random variable in computing an upperbound to the probability of events ${\cal E}_1,\,{\cal E}_2$. A second problem is that in the
deterministic case we assumed that the $\ell_2$ norm of each column, $\bg_j$ is deterministically normalized to unity. In the Gaussian case only the expected power is normalized
to unity. Note also that for the output noise model considered in this paper $\bsg = \mathbf{I}$. Therefore $\lambda_{\min}(\bsg^{-1}) = 1$. Following along the lines of the
proof of Lemma~\ref{lem:E1} we see that,
\begin{align*}
\prob({\cal E}_1 \mid \bG) \leq \exp \left\{- \frac{\sigma_{\bG,\min}^{2} \beta^2 \snr}{32 \max_j\|\bg_j\|_2}\right\} \exp\{\log 2n\}
\end{align*}
We need to now characterize a lower bound for ${\sigma_{\bG,\min}^2\over \max_j\|\bg_j\|_2}$.
To this end we observe that,
\begin{align} \label{eq:lb}
\Pr \left ({\sigma_{G,\min}^2 \over \max_j \|\bg_j\|_2}\geq {(1-\eta)^2\over 1+\epsilon}\right ) \geq \Pr(\sigma_{G,\min} \geq& (1-\eta)^2,\, \max_j \|\bg_j\|_2 \leq 1+\epsilon)\\ &\geq 1-(\Pr(\sigma_{G,\min} \leq (1-\eta)^2)+\Pr(\max_j \|\bg_j\|_2 \geq 1+\epsilon)) \nonumber
\end{align}
This implies that we should characterize $\sigma_{G,\min}$ and $\max_j \|\bg_j\|_2$ separately. We appeal to the following lemma in \cite{CandesIT05}, to characterize $\sigma_{G,\min}$.
\begin{lem}
\label{lem:rip} Suppose the sparsity is $\alpha_n = k/n$ and we consider a function $f(q) := \sqrt{n/m}\left (\sqrt{q} + \sqrt{2H_2(q)}\right )$, where $H_2(q) := -q \log q -
(1-q) \log(1-q)$. Let $\bG$ be an $m\times n$ matrix drawn from a Gaussian ensemble with $g_{ij}\stackrel{d}{\sim}{\cal N}(0,1/m)$. Then it follows that $\sigma_{\bG,\min}$
described in Equation~\ref{eq:rip} has the following concentration property,
\begin{align}
\prob \left ( \sigma_{\bG,\min} \leq 1-\eta \right ) \leq 2 \exp \left ( -{n\epsilon H_2(\alpha_n)\over 2} \right )\stackrel{\Delta}{=}\delta_1(n,\alpha_n,\epsilon)
\end{align}
where, $\eta=2 (1 + \eps)f(2 \alpha) + (1 + \eps)^2 f^2(2\alpha)$.
\end{lem}
We consider the following concentration result to characterize maximum power of the columns of $\bG$.
\begin{lem}
\label{lem:norm_g} Let $\bG$ be drawn from an IID Gaussian ensemble with $g_{ij}\stackrel{d}{\sim} {\cal N}(0,1/m)$. Let $\bg_j,\,j=1,\,2\,\ldots,\,n$ be the columns of $\bG$. Then, for any $\epsilon > 0$, it follows that,
\[
\prob(\max_j \|\bg_j\|^2_2 \geq 1+\epsilon)\leq \exp \left( - {m\over 2} (\log(1 + \epsilon) + \epsilon) + \log n \right)\stackrel{\Delta}{=}\delta_2(m,n,\epsilon)
\]
\end{lem}
\begin{proof}
Clearly $X:=m\|\bg\|^2_2$ is $\chi^2$ distributed with degree $m$ and its moment generating function is $\mathbb{E}(e^{tX})=(1-2t)^{-m/2}$. From Chernoff bound,
\begin{eqnarray*}
\Pr(X\geq a)\leq \frac{\mathbb{E}(e^{tX})}{e^{ta}}=\frac{(1-2t)^{-m/2}}{e^{ta}}
\end{eqnarray*}
Choosing $a = m(1+\epsilon)$ and $t=\frac{1}{2}(1-m/a)=\frac{\epsilon}{2(1+\epsilon)}$, we have
\[\Pr(\|\bg\|^2_2 \geq 1+\epsilon)\leq \exp \left( - {m\over 2} (\log(1 + \epsilon) + \epsilon) \right)\]
The proof then follows by employing the union bound.
\end{proof}
Putting Lemmas~\ref{lem:rip} and \ref{lem:norm_g} together with Equation~(\ref{eq:lb}) and taking the expectation with respect to $\bG$ we get,
\begin{align*}
\prob({\cal E}_1) =& E_{\bG} \left ( \prob({\cal E}_1 \mid \bG) I_{\Gamma} + \prob({\cal E}_1 \mid \bG) I_{\Gamma^c}\right ) \\ &\leq \exp \left\{- \frac{(1-\eta)^2 \beta^2 \snr}{32 (1+\eps)}\right\} \exp\{\log 2n\}(1-\delta) + \delta
\end{align*}
where $\Gamma= \{ \bG: {\sigma_{\bG,\min}\over \max_j \|\bg_j\|_2} \leq {(1-\eta)^2 \over (1+\eps)}\}$ and
$\delta = \delta_1(n,\alpha_n,\epsilon)+\delta_2(m,n,\epsilon)$. Note that $\prob(\Gamma^c) \leq \delta$ and $\delta$ can be made arbitrarily small for $m =\Omega(\log(n))$ and $k$ sufficiently large. We are now left to ensure that the first term in the RHS of the above equation can be made small as well. For this purpose we need
\begin{align}
\label{eq:supp_CS_step1} \frac{(1-\eta)^2 \beta^2 \snr}{(1 + \epsilon) 32} = (1+\gamma) \log 2 n
\end{align}
for some arbitrary $\gamma > 0$. Let $\eta_1 = \left(\frac{32(1+ \gamma)(1 + \epsilon) \log 2n}{\beta^2 \snr}\right)^{1/2}$. This implies that it is sufficient that,
\begin{align}
1 - \eta\geq \eta_1 \implies \eta \leq
1 - \eta_1\\
\implies (1 + \eps)f(2\alpha)(2 + (1 + \eps) f(2\alpha)) + 1 \leq 1 + (1 - \eta_1)\\
\implies (1 + (1 + \eps) f(2\alpha))^2 \leq 2 - \eta_1 \implies (1 + \eps) f(2\alpha) \leq \sqrt{2 - \eta_1} - 1
\end{align}
For this inequality to be satisfied we need $\eta_1 \leq 1$. A sufficient condition for support recovery can be obtained by substituting for $\eta$ and we get
$$
\eta_1 ={32(1+\gamma)(1+\epsilon)\log(2n)\over \beta^2 \snr} < 1,\,\,{n\over m} \leq \frac{1}{(1 + \eps)^2(\sqrt{2\alpha} + \sqrt{2H_2(2 \alpha)})^2}(\sqrt{2-\eta_1}-1)^2
$$
Since $(\sqrt{2\alpha} + \sqrt{2H_2(2 \alpha)})^2 \leq 6 H_2(2 \alpha)$ and $\gamma,\,\eps$ can be made arbitrarily small, the result now follows for event ${\cal E}_1$. %\proofover

We are now left to bound the probability of event ${\cal E}_2$. This case is simple since the normalizing factor $\max_j \|\bg_j\|_2$ is no longer relevant as seen from the proof of Lemma~\ref{lem:E2}. It suffices to ensure that $\sigma_{\bG,\min}$ needs to be bounded away from zero. However, note that we already have this from bounding the probability of event ${\cal E}_1$. The result now follows.
\section{Recovery for Arbitrary Distortions: Bayesian signal model}
\label{sec:average_dist}
In this section we switch to a Bayesian signal model from the worst-case setting considered in the previous section. There are a number of reasons for considering such a model: \\
({\bf A}) For both the input and output noise models we need the $\snr$ to scale as $\Omega(\log(n))$ for exact support recovery regardless of the number of measurements.\\
({\bf B}) For exact support recovery in the worst-case setup we require that the minimum singular values of all sub-matrices of $\bG$ as described in Equation~(\ref{eq:rip}) be uniformly bounded away from zero (Theorem~\ref{thm:mainsuppthmsuff1}). This arises because a worst-case signal, $\bX$, matched to the smallest singular value can be chosen. However, this problem may not arise in the average case setting. \\
({\bf C}) The situation is worse for the input noise model. Even with $\snr$ of $\Omega(\log(n))$ the number of measurements required is linearly proportional to signal dimension. \\
({\bf D}) Theorem~\ref{thm:dist_supp} points out that even with distortion we can only hope to reduce the $\snr$ but not the number of measurements.

Consequently, it is worth exploring whether these results can be improved in the average Bayesian case.  Fundamentally, the idea is that if we remove a sufficiently small set of
signals then it is conceivable that the results could be more promising. %Clearly, one consequence of such a direction is that we may no longer need $\sigma_{\bG,\min}$ as in
%Theorem~\ref{thm:mainsuppthmsuff1} to be uniformly bound away from zero.

In the following we first develop novel lower and upper bounds to probability of error subject to a distortion in reconstruction. The main ingredient in realizing these bounds
is the use of the minimal covering property of the rate distortion function.  We begin with a minimal cover as a functional mapping of the source to the set of rate distortion
quantization points. Then for the lower bound to the probability of error we follow the steps of the proof Fano's inequality, \cite{Cover} which we appropriately modify to
address \emph{detection} of the correct quantization point corresponding to the true $\bX$. Similarly for the upper bound to the probability of error we propose a minimum
distance decoder (ML decoder for AWGN noise) over the set of rate distortion quantization points and derive a closed form result for the particular case of $\ell_2$ distortion.

\subsection{Lower bound- modified Fano's inequality}
\label{sec:pelowerbound}

In the following we will use $\bX$ and $X^n$ interchangeably. The main reason for introducing this notation is that we will deal with $n$-dimensional probability distributions
over $\bX$ induced by the product measure $P_{X^n} = P_X\times...\times P_X (n \,\,\mbox{times})$.
\begin{lem}
\label{lem:lowerbound} Given observation(s) $\bY$ for the sequence $X^n \triangleq \left\{X_1,\,\ldots,\,X_n\right\}$ of random variables drawn IID with $X_i
\stackrel{d}{\sim}P_X$. Let $\hat{X}^n(\bY)$ be the reconstruction of $X^n$ from $\bY$. Let the distortion measure be given by $d(X^n,\hat{X}^n(\bY)) = \sum_{i=1}^{n}
d(X_i,\hat{X}_i(\bY))$.  Then given $\epsilon >0$ for sufficiently large $n$ we have
\begin{align*}
\prob \left( \frac{1}{n}d(\hat{X}^n(\bY),X^n)  \geq d_0\right) \geq \dfrac{R_{X}(d_0)  - K(d_0,n) - \frac{1}{n}\ind(X^n;\bY)}{R_{X}(d_0) + \epsilon } + \epsilon
\end{align*}
where $K(d_0,n)$ is the logarithm of the number of neighbors of a quantization point in the n-dimensional rate-distortion mapping) and $ R_{X}(d_0)$ is the corresponding
(scalar) rate distortion function for $X$.
\end{lem}

We have the following result for the special case of finite alphabets with Hamming distortion.
%\subsubsection{Discrete ${\cal X}$ under Hamming Distortion}

\begin{lem}
\label{lem:discrete_LB}Given observation(s) $\bY$ for the sequence $X^n \triangleq \left\{X_1,...,X_n\right\}$ of random variables drawn i.i.d. according to $P_X$ and $X_i \in
{\cal X},\,\, |{\cal X}| < \infty$. Let $\hat{X}^n(\bY)$ be the reconstruction of $X^n$ from $\bY$.  For hamming distortion $d_{H}(\cdot,\cdot)$ and for distortion levels,
\begin{align*}d_0 \leq \min \left\{1/2, (|{\cal
X}| - 1)\min_{x\in {\cal X}} P_X(x)\right\}\end{align*} we have
\begin{align*}  \prob\left( \frac{1}{n}d_{H}(X^n,\hat{X}^n(\bY)) \geq d_0\right)
 \geq \dfrac{nR_{X}(d_0) - \ind(X^n;\bY) - 1 - \log nd_0}{n
\log(|{\cal X}|) - n \left(H_2(d_0) + d_0 \log (|{\cal X}|-1) + \frac{\log nd_0}{n}\right)}
\end{align*}
\end{lem}

%\begin{proof} See Appendix. \end{proof}

\subsection{Constructive upper bound to probability of error for $\ell_2$ distortion} \label{sec:peupperbound}

In this section we will provide a constructive upper bound to the probability of error in reconstruction subject to an average squared distortion level for the output noise
model. To this end assume that we are given a \emph{minimal $d_0$ cover} as described in Theorem \ref{thm:mincover} of \cite{Kontoyiannis01}. Specifically, we have a set of balls, ${\cal B}_i \subset \Real^n,\,i=1,\,2\,\ldots,2^{n(R_{X}(d_0) + \epsilon)}$, of diameter $2\sqrt{nd_0}$ such that, for any $\epsilon >0$ we have for sufficiently large $n$ that,
$$
\Pr\{\bigcup_{i=1}^{N_{\epsilon}(n,d_0)}{\cal B}_i\} \geq 1-\epsilon
$$
where $R_{X}(d_0)$ is the (scalar) rate distortion function for $X \stackrel{d}{\sim} \prob_X$ and $N_{\epsilon}(n,d_0)=2^{n(R_{X}(d_0) + \epsilon)}$.
Each ball ${\cal B}_i$ is represented by a quantization points $\bZ_i \doteq
Z_{i}^{n}$. Thus with high probability for any $\bX$ there exists a point, $\bZ_i$ to which it can be mapped to such that the distortion is less than $d_0$.

We consider a modified maximum likelihood estimator to establish an achievable upper bound. Given $\bG$ and the rate distortion points $\bZ_{i}$, we enumerate the set of points, $\bG \bZ_{i} \in \Real^{m}$. Then given the observation $\bY$ we map it to the nearest point $\bG \bZ_{i} \in \Real^{m\times 1}$. Our estimator $\hat \bX(\bY)$ then outputs $\bZ_i$. We refer to Figure
\ref{fig:nearest_decod} for an illustration.

%prove the result by converting the problem to a Max-Likelihood (ML) \emph{detection} set-up over the set of rate-distortion quantization points given by the minimal cover as
%follows. Given $\bG$ and the rate distortion points $\bZ_{i}$ corresponding to the functional mapping $f(\bX)$, we enumerate the set of points, $\bG \bZ_{i} \in \Real^{m}$. Then
%given the observation $\bY$ we map $\bY$ to the nearest point $\bG \bZ_{i} \in \Real^{m\times 1}$ and is decoded as $\bZ_i$.

\begin{lem}
\label{lem:PeUpper} Given observation $\bY = \bG \bX + \frac{\bN}{\snr}$ for the sequence $\bX \doteq X^n \triangleq \left\{X_1,\,\ldots,\,X_n\right\}$ of random variables drawn
IID with $X_i \stackrel{d}{\sim}P_X$. Let $\hat{X}^n(\bY)$ be the reconstruction of $X^n$ from $\bY$. Then for any $\epsilon >0$ we have for sufficiently large $n$,
\begin{align} \label{eq:PeUpper}
\prob(\|\hat{\bX}(\bY) - \bX\|^2 \geq 2n d_0) \leq (1-\epsilon)\exp\left\{- \frac{\snr \|\bG(\bZ_i - \bZ_j)\|^2}{32} \right\} 2^{nR_{X}(d_0)} + \epsilon
\end{align}
where $\bZ_i$ and $\bZ_j$ are any two quantization points such that $\|\bZ_i - \bZ_j\| = 4\sqrt{n d_0} $.
\end{lem}

\begin{proof}
To compute the probability of error we first consider a pairwise error probability, namely,
\begin{align}
\label{eq:pe_ub_rate_dist} \prob_e(i,j) =  \prob \left\{\bN: \bX \in {\cal B}_i \rightarrow \bZ_j \mid d({\cal B}_i,{\cal B}_j) \geq 2n d_0 ,\bG \right\}
\end{align}
where, $d({\cal B}_i,{\cal B}_j)$ is the minimum squared distance between any two points, $\bX_i \in {\cal B}_i$ and $\bX_j \in {\cal B}_j$.
Under the minimum distance estimator we have,
\begin{align}
\prob_e(i,j) = \prob \left\{ \bN: \|\bG \bX + \frac{\bN}{\sqrt{\snr}} - \bG \bZ_i\|^2 \geq \|\bG \bX + \frac{\bN}{\sqrt{\snr}} - \bG \bZ_j\|^2\right\}
\end{align}
where we have omitted the conditioning variables and equations for brevity. Simplifying the expression inside the probability of error we get that,
\begin{align} \prob_e(i,j) = \prob \left\{ 2 \frac{\bN^T}{\sqrt{\snr}} \frac{\bG (\bZ_j - \bZ_i)}
{\|\bG (\bZ_j - \bZ_i)\|} \geq \frac{\|\bG(\bX -\bZ_j)\|^2 - \|\bG(\bX-\bZ_i)\|^2}{\|\bG(\bZ_j-\bZ_i)\|}\right\}
\end{align}
In other words we are asking for the pairwise probability of error in mapping a signal that belongs to the distortion ball ${\cal B}_i$ to the quantization point $\bZ_j$ of the
distortion ball ${\cal B}_j$ under the noisy mapping $\bG \bX + \bN$ such that the set (squared) distance between the distortion balls is $\geq 2 n d_0$, see Figure
\ref{fig:nearest_decod}.

Under the assumption that the noise $\bN$ is an AWGN noise with unit power in each dimension, its projection $N$ onto the unit vector $ \frac{\bG (\bZ_j - \bZ_i)}{\|\bG(\bZ_j - \bZ_i)\|}$ is also AWGN with unit power. Thus we have
\begin{align*}
\prob_e(i,j)  & = \prob \left\{ \frac{N}{\sqrt{\snr}}  \geq \frac{\|\bG(\bX -\bZ_j)\|^2 - \|\bG(\bX-\bZ_i)\|^2}{2 \|\bG(\bZ_j-\bZ_i)\|}\right\} \\
 & \leq   \prob \left\{ \frac{N}{\sqrt{\snr}}  \geq  \min_{\bX \in {\cal B}_i}\frac{\|\bG(\bX -\bZ_j)\|^2 - \|\bG(\bX-\bZ_i)\|^2}{2 \|\bG(\bZ_j-\bZ_i)\|}\right\}
\end{align*}
where we have further upper bounded the probability of the pairwise error via choosing the worst case $\bX$ that minimizes the distance between the ball ${\cal B}_i$ and the
quantization point $\bZ_j$ and maximizes the distance from the quantization point $\bZ_i$ within the distortion ball ${\cal B}_i$.

\begin{figure}
\begin{centering}
\includegraphics[width = 4 in]{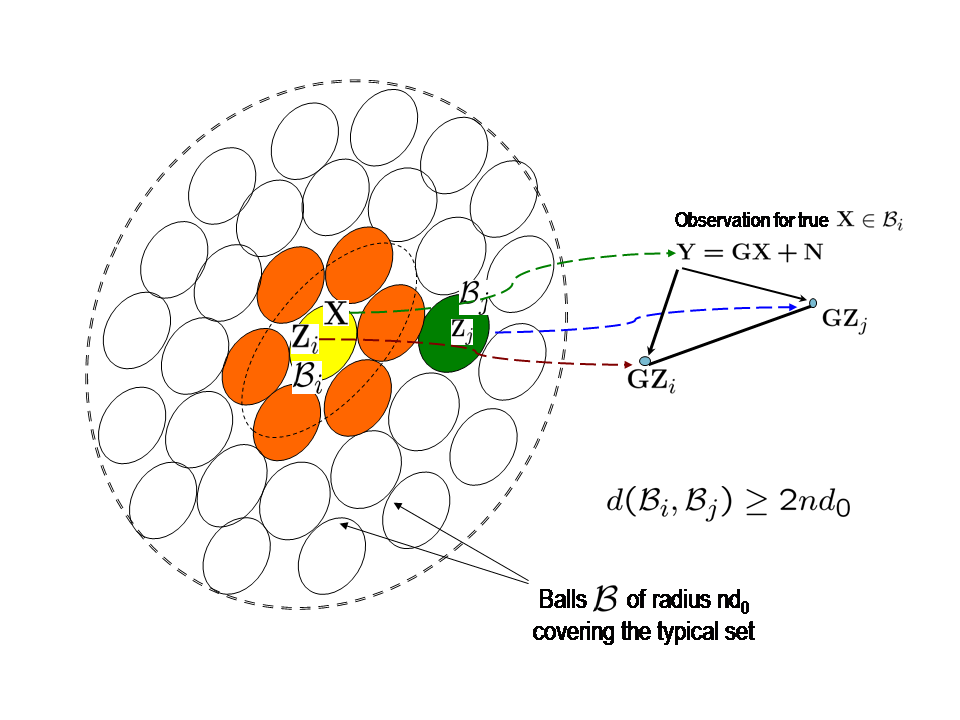}
\caption{\small Figure showing the rate distortion cover by balls ${\cal B}$ of radius $\sqrt{nd_0}$. The ML decoding over the set of rate distortion quantization points (identified as centers of the distortion balls) consists of mapping $\bY$ to the correct distortion ball for $\bX$ using a minimum distance decoder. Shown in the figure is a pair-wise error event for mapping $\bX \in {\cal B}_i$ to quantization point $\bZ_j \in {\cal B}_j$ that is at a set distance of $2nd_0$ from ${\cal B}_i$ to which $\bX$ belongs.} \label{fig:nearest_decod}
\end{centering}
\end{figure}

For the case of squared distortion and covering via spheres of average radius $d_0$, it turns out that the worst case $\bX$ is given by $\bX = \frac{3\bZ_i + \bZ_j}{4}$ and
$\|\bZ_i - \bZ_j\| = 4\sqrt{n d_0}$. Plugging this value in the expression we have for the worst case pairwise probability of error that
\begin{align*}
\prob_e(i,j)  \leq   \prob \left\{ N  \geq  \frac{\sqrt{\snr} \|\bG(\bZ_i -\bZ_j)\|}{4}\right\}
 \leq \exp\left\{- \frac{\snr \|\bG(\bZ_i -\bZ_j)\|^2}{32 }\right\}
\end{align*}
where the second inequality follows by the standard upper bound to the error function. Now we apply the union bound over the set of rate distortion quantization points $\bZ_j$
minus the set of points that are the neighbors of $\bZ_i$ (see figure \ref{fig:nearest_decod}). The maximum number of such points is given by $N_{\epsilon}(n,d_0)= 2^{n(R_{X}(d_0)+\epsilon)}$, where $R_{X}(d_0)$ is the scalar rate distortion function, \cite{Cover}. Hence we have,
\begin{align*}
\prob(\|\hat{\bX} - \bX\|^2 \geq 2n d_0 \mid \bX \in \bigcup_i {\cal B}_i) \leq \exp\left\{- \frac{\snr \|\bG(\bZ_i - \bZ_j)\|^2}{32} \right\} 2^{n(R_{X}(d_0)+\epsilon)}
\end{align*}
with $\|\bZ_i - \bZ_j\| = 4\sqrt{n d_0}$. To finish the proof we note that with probability $(1-\epsilon)$, the signal $\bX$ belongs to one of the balls ${\cal B}_j$. Thus taking expectations with respect to $\bX$ the result follows.
\end{proof}
\section{Approximate Recovery: Bayesian Bounds}
\label{sec:scap_bounds}

In this paper we will consider the following mixture model for explicit evaluation of the bounds.
\begin{align}
\label{eq:source_random} \bX_i \stackrel{d}{\sim} P_X = \alpha {\cal N}(\mu_1,\sigma_{1}^2) + (1 - \alpha) {\cal N}(\mu_0,\sigma_{0}^{2})
\end{align}
i.e., each component $X_i$ of $\bX$ is IID $P_X$ defined above. It is easy to see that for $\mu_1 = 1$, $\mu_0 = 0$ for $\sigma_0 = 0$ this mixture model for large enough $n$
results in an approximately $k = \alpha n$ sparse sequence. We use $\sigma_1 = 0$ to model a binary discrete case and $\sigma_1 = 1$ to model a continuous valued case. It is
worth pointing out that this model has been used previously in several papers, e.g. see \cite{Sarvotham_TR06,Fletcher06} to probabilistically model sparse signals.

\subsection{Discrete $\bX$: Support recovery}
It is easy to see that using a binary signal model for $\bX$ one can address the support recovery problem in the Bayesian setting. Under this case $\bX$ is drawn IID according
to,
\begin{align}
\label{eq:dist_discrete} P_{X} = \alpha \delta(X-\beta) + (1 - \alpha) \delta(X)\,,: \,\,\alpha \leq 0.5
\end{align}
where, $\delta(\cdot)$ is the usual singular measure.
Note that it follows from Asymptotic Equipartition Property (AEP), see \cite{Cover}, that  asymptotically the $n$-dimensional probability distribution uniformly concentrates on
the set of exactly $k$-sparse sequences $\Xi_{\{0,\beta\}}^{\alpha n}$, i.e. given $\epsilon > 0, \exists n$ such that $P_{X^n}\left(\Xi_{\{0,\beta\}}^{\alpha n}\right) \geq 1 -
\epsilon$. Thus these bounds can be compared to the worst-case setup of Section~\ref{sec:scap_support} when $\bX \in \Xi_{\{0, \beta\}}^{k}\, , k = \alpha n $. For this discrete case we have the following main
results stated in terms of the scalar rate distortion function $R_X(d_0)$ with Hamming distance as the distortion measure. Note that for this  case $R_{X}(d_0) = H_2(\alpha) -
H_2(d_0): \, d_0 \leq \alpha$.

\begin{thm}
\label{thm:scap_binom} Consider the input noise model of Equation~(\ref{eq:model_snr1}) and the binary model for $\bX$ as described above. Then,
\begin{itemize}
\item[a.] \textbf{Necessity}: Asymptotically as $n\rightarrow \infty$ if $m \leq \dfrac{n R_X(d_0)}{0.5 \log (1 + \alpha \beta^2 \snr)}$ there does not exist any algorithm that
recovers the signal to within an average Hamming distortion of $d_0$.
\item[b.] \textbf{Sufficiency}: Asymptotically as $n\rightarrow \infty$, it is sufficient that $m \geq \dfrac{n R_X(d_0/2)}{0.5 \log (1 + \frac{d_0 \beta^2 \snr}{2})}$ for the constructive ML estimator of section
\ref{sec:peupperbound} to reliably recover the signal to within Hamming distortion of $d_0$.
\end{itemize}
\end{thm}

\begin{proof}
To prove part (a) note that from Lemma \ref{lem:discrete_LB} for the probability of error to approach zero implies that the numerator in the lower bound approach zero. This
implies that we need,
\begin{align}
\frac{n}{m} \leq \dfrac{\frac{1}{m} \ind(\bX;\bY|\bG)}{R_{X}(d_0) - \frac{1}{n} - \frac{\log nd_0}{n}}
\end{align}
To this end recall that $\bY = \bG(\bX + \frac{1}{\sqrt{\snr}}\bN)$. Consider the SVD of $\bG = \bU \bS \bV^*$, where $\bU,\,\bV$ are  orthonormal matrices and $\bS = [
\mathbf{D}\,\,\, \mathbf{0} ]$, with $\mathbf{D}$ a positive diagonal matrix. From \cite{Verdu_Book04} it follows that $\bU,\,\bS,\,\bV$ are independent random matrices.
Furthermore $\bU$ and $\bV$ are isotropically random. By linearly transforming $\bY$ by pre-multiplying by $\bD^{-1} \bU^*$ we get an equivalent system of equations with
\begin{align} \label{e.svd}
\tilde \bY = \bV_1^*\bX + \frac{1}{\sqrt{\snr}}\bV_1^*\bN
\end{align}
where $\bV_1^*$ is the matrix formed from the first $m$ rows of $\bV^*$. Now note that since the rows of $\bV_1^*$ are orthogonal and normalized  $\tilde \bN =
\frac{1}{\sqrt{\snr}}\bV_1^*\bN$ is IID Gaussian with each component having zero mean and variance $1/SNR$. This transformation implies that $\ind(\bX;\bY|\bG) =
\ind(\tilde{\bY};\tilde{\bX}|\bV_1)$ since $\bV$ is independent of $\bU$ and $\bS$. Now by direct computation it follows that,
$$
\ex_{\bV} \ind(\tilde \bY;\tilde \bX \mid \bV_1) \leq h(\tilde \bY \mid \bV_1) - h(\tilde \bY \mid \bV,\tilde \bX) \leq \frac{m}{2} \log(1 + \snr \alpha \beta^2)
$$
where to get the last inequality we have used the fact that $h(\tilde \bY \mid \bV,\tilde \bX)$ is the entropy of noise $\tilde \bN$ and  for the first term, $h(\tilde \bY \mid
\bV_1)$, we have used the fact that a Gaussian distribution maximizes the entropy over all other random variables with zero mean and identical variance~\cite{Cover}. Finally,
for sufficiently large $n$ the term $\frac{1}{n} + \frac{\log nd_0}{n}$ can be made arbitrarily small and the result follows.

We will now prove part (b). In order to simplify the derivation we again focus on Equation~(\ref{e.svd}). Following the proof of Lemma \ref{lem:PeUpper} the pairwise error can
now be computed as follows
\begin{align}
\prob_e(i,j) & \leq   \prob \left\{ N  \geq  \frac{\sqrt{\snr} \|\bV_1^{*}(\bZ_i -\bZ_j)\|}{4} \left | \right . \bV_1 \right\}
 \leq \exp\left\{- \frac{\snr \|\bV_1^{*}(\bZ_i -\bZ_j)\|^2}{32 } \right\}
\end{align}
To compute the error probability we will need to take the expectation over $\bV_1$ and apply the union bound to bound the error probability over all error patterns. To simplify
the expectation over $\bV_1$ we let,
\begin{align}
\phi(\bD,\bV_1) = \exp\left\{- \frac{\snr \|\bD \bV_1^{*}(\bZ_i -\bZ_j)\|^2}{32 } \right\}
\end{align}
where, $\bD$ is a positive diagonal random matrix independent of $\bV_1^*$. Note that our  problem reduces to bounding expectation of $\phi(I_m,\bV_1)$ over $\bV_1$. Note that
when $\sigma_{\max}(\bD) \leq 1$ we have $\phi(I_m,\bV_1) \leq \phi(\bD,\bV_1)$. Next, note that trivially we have,
\begin{align}
\phi(I_m,\bV_1)) I_{\{\sigma_{\max}(\bD) \leq 1\}} \leq \phi(\bD,\bV_1) I_{\{\sigma_{\max}(\bD) \leq 1\}} + \phi(\bD,\bV_1) I_{\{\sigma_{\max}(\bD) \geq 1\}}
\end{align}
where $I_{\{\cdot\}}$ denotes the indicator function. Consequently, we can take expectations over the two independent matrices $\bD$ and $\bV_1$ to obtain,
\begin{align}
E_{\bV_1}(\phi(I_m,\bV_1))) Prob(\sigma_{\max}(\bD)\leq 1) \leq E_{\bD,\bV_1} \exp\left\{- \frac{\snr \|\bD \bV_1^{*}(\bZ_i -\bZ_j)\|^2}{32 } \right\}
\end{align}
Note that we can introduce a isotropically random unitary matrix $\bU$,  namely, $\exp\left\{- \frac{\snr}{32} \|\bD \bV_1^{*}(\bZ_i -\bZ_j)\|^2 \right\}=\exp\left\{-
\frac{\snr}{32} \|\bU\bD \bV_1^{*}(\bZ_i -\bZ_j)\|^2 \right\}$ without modifying the result. Now the matrix $\bH = \bU\bD \bV_1^*$ can be identified by a suitable IID Gaussian
matrix when $\bU,\,\bD,\,\bV$ are chosen independently and $\bU$ and $\bV$ are chosen uniformly from set of all unitary matrices; the positive diagonal matrix $\bD$ is
distributed according to the distribution of singular values of a Gaussian matrix. To ensure a tight approximation we need to choose a Gaussian matrix such that
$\prob(\sigma_{\max}(\bD) \leq 1)$ approaches one. This can be accomplished by choosing $\bH$ as an IID Gaussian ensemble with each component $h_{ij} \stackrel{d}{\sim} {\cal
N}(0,{1 \over (1+\sqrt{m/n})n})$. Then following similar steps as in the proof of Lemma \ref{lem:PeUpper} we arrive at a similar upper bound,
\begin{align*}
\prob(\|\hat{\bX} - \bX\|^2 \geq 2n d_0) \leq (1 - \epsilon) \exp\left\{- \frac{\snr \|\bH(\bZ_i - \bZ_j)\|^2}{32} \right\} 2^{n(R_{X}(d_0)+\epsilon)} + \epsilon
\end{align*}
where, $\|\bZ_i - \bZ_j\| = 4\sqrt{n d_0}$. Since $\epsilon$ is arbitrary, the result then follows by taking expectation with respect to $\bH$ and using the moment generating
function of the $\chi^2$ random variable, \cite{Vahid}.
\end{proof}

\begin{thm}
\label{thm:scap_binom_CS}Consider the output noise model of Equation~(\ref{eq:model_snr}) and the binary model for $\bX$ as described above. Then,
\begin{itemize}
\item[a.] \textbf{Necessity}: Asymptotically as $n\rightarrow \infty$ if $m \leq \dfrac{n R_X(d_0)}{0.5 \log (1 + \frac{n}{m} \alpha \beta^2 \snr)}$ there does not exist any algorithm that recovers the signal to within an average Hamming distortion of
$d_0$.
\item[b.] \textbf{Sufficiency}: Asymptotically as $n\rightarrow \infty$ it is sufficient that $m \geq \dfrac{n R_X(d_0/2)}{0.5 \log (1 + \frac{n}{m} \frac{d_0 \beta^2 \snr}{2})}$ for the constructive ML estimator of section
\ref{sec:peupperbound} to reliably recover the signal to within Hamming distortion of $d_0$.
\end{itemize}
\end{thm}

\begin{proof}
The proof of part (a) follows along the same lines as that of \ref{thm:scap_binom} with the following modification to the upper bound of the mutual information expression,
\begin{align}
\ex_{\bG} \ind(\bX;\bY|\bG) \leq \frac{m}{2} \log(1 + \frac{n \alpha \beta^2 \snr }{m})
\end{align}
The proof of part (b) follows from the upper bound to the probability of error in Lemma \ref{lem:PeUpper} by taking expectation with respect to $\bG$ and using the moment
generating function of the $\chi^2$ random variable, see \cite{Vahid}.
\end{proof}
We will now reduce the implicit expression in the above Lemma to derive some explicit conditions on the number of measurements $m$. To this end we have the following corollary.

\begin{cor} \label{cor:num_meas_discrete} Consider the output noise model of Equation~(\ref{eq:model_snr}) and the binary model for $\bX$ as described above. Then,
(a) Asymptotically as $n\rightarrow \infty$ if  $\snr \leq \frac{2 R_X(d_0)}{\alpha \beta^2}$ and $m\leq 2 n R_X(d_0)$ there exists no algorithm that can recover $\bX$  to
within an average Hamming distortion of $d_0$; (b) On the other hand asymptotically as $n\rightarrow \infty$ it is sufficient that $\snr \geq \frac{200 R_X(d_0/2)}{d_0 \beta^2}$
with $m \geq 2.08 nR_X(d_0/2)$ for the constructive ML estimator of section \ref{sec:peupperbound} to recover $\bX$ to within an average Hamming distortion of $d_0$.
\end{cor}
\begin{proof}
To begin with we will focus on the sufficient conditions. Denote by $c = \frac{nR_{X}(d_0/2)}{m}$. Also let $\eta = \frac{d_0 \beta^2 \snr}{2 R_X(d_0/2)} $. Then from part (b)
of Theorem \ref{thm:scap_binom_CS} we have as a sufficient condition that,
\begin{align}
f(c) = 0.5 \log ( 1 + c \eta)  - c \geq 0
\end{align}
In particular we want to find $\max\{c | f(c) \geq 0\}$. To this end note that $f(c) = 0 $ at $c = 0$. Also for there to exist any positive $c$ such that $f(c) > 0$ it is
required that $\eta \geq 2$. In particular $\eta \geq 2$ is the condition for a positive derivative near zero. This implies that $ \frac{d_0 \beta^2 \snr}{2 R_X(d_0/2)} \geq 2$
or $\snr \geq \frac{4 R_X(d_0/2)}{d_0 \beta^2}$. Given that this condition is satisfied, $c = \frac{1 - 2/\eta}{2}$ lies in the \emph{feasible} region. Therefore $m = \frac{2}{1
- 2/\eta} n R_{x}(d_0/2)\approx 2 n R_X(d_0/2)$ is a sufficient condition for reliable recovery for some sufficiently large $\eta> 2$, i.e. for $\snr \geq \frac{4
R_X(d_0/2)}{d_0 \beta^2}$ . In particular if we choose $\eta = 100$ then $\snr \geq \frac{200 R_X(d_0/2)}{d_0 \beta^2}$ and $m \geq 2.08nR_X(d_0/2)$ is sufficient for reliable
recovery.

Analyzing part (a) of the Theorem \ref{thm:scap_binom_CS} in a similar manner, one can show that if $\snr \leq \frac{2 R_X(d_0)}{\alpha \beta^2}$ and $m \leq 2 n R_{X}(d_0)$
there exits no algorithm that can reliably recover $\bX$ to within the desired distortion level.
\end{proof}

\begin{remark} One immediate observation from the above analysis is that unlike the worst case set-up one can indeed
tradeoff the number of measurements with distortion in the Bayesian set-up.
\end{remark}

\subsection{Continuous $\bX$: $\ell_2$ recovery}

Under this case $\bX$ is drawn IID according to,
\begin{align}
\label{eq:dist_cont} P_{X} = \alpha {\cal N}(0,\beta^2) + (1 - \alpha) \delta(X)
\end{align}
For this case we have the following main results. The results are stated in terms of the scalar rate distortion  function $R_X(d_0)$ given by $R_X(d_0) = H_2(\alpha) +
\frac{\alpha}{2} \log \frac{\alpha}{d_0}: \, d_0 < \alpha$, (see section \ref{subsec:ratedist_gauss} for the derivation of this result). Notice in the following that in contrast
to the discrete case where $d_0 \leq \alpha$ here we impose $d_0 \leq \alpha/2$ and for reasonable reconstruction one typically desires $d_0 = \epsilon \alpha$ for some small
$\epsilon > 0$. The reason that we require $d_0 \leq \frac{\alpha}{2}$ is due to the additional term of $K(n,d_0)$ in the modified Fano's inequality~\ref{lem:lowerbound} which appears in the
continuous setting.

\begin{thm}
\label{thm:scap_continuous} Consider the input noise model of Equation~(\ref{eq:model_snr1}) and the mixture model for $\bX$ as described above. Then,
\begin{itemize}
\item[a.] \textbf{Necessity}: Asymptotically as $n\rightarrow \infty$ if $m \leq \dfrac{n (R_X(d_0)- \frac{\alpha}{2} \log 2)}{0.5 \log (1 + \alpha \beta^2 \snr)}$ there does not exist any algorithm that recovers the signal to
within an average $\ell_2$ distortion of $d_0$.
\item[b.] \textbf{Sufficiency}: Asymptotically as $n\rightarrow \infty$ it is sufficient that $m \geq \dfrac{n R_X(d_0/2) }{0.5 \log (1 + \frac{d_0 \beta^2 \snr}{2})}$ for the
constructive ML estimator of section \ref{sec:peupperbound} to reliably recover the signal to within an average $\ell_2$ distortion of $d_0$.
\end{itemize}
\end{thm}

\begin{proof}
For part (a) first note that from Theorem \ref{thm:scap_binom} we have $\ex_{\bG} \ind(\bX;\bY|\bG) \leq \frac{m}{2} \log(1 + \beta^2 \alpha \snr)$. From Lemma \ref{lem:lowerbound} it follows that for feasibility of recovery to with distortion $d_0$ (asymptotically) it is
required that,
\begin{align}
\frac{n}{m} \leq \frac{\frac{1}{m}\ex_{\bG} \ind(\bX;\bY|\bG)}{R_{X}(d_0) -K(d_0,n)}
\end{align}
The result then follows by noting that $ |K(d_0,n) - 0.5 \alpha \log 2| < \epsilon$ with $\epsilon$ arbitrarily small for large enough $n$, see e.g. \cite{zegerIT94}. Note that
for the case at hand in order for the expression $R_{X}(d_0) -K(d_0,n)$ to remain positive and hence meaningful, $d_0 \leq \alpha/2$. The proof of part (b) follows exactly along
the same lines as the proof of part (b) in Theorem \ref{thm:scap_binom}.
\end{proof}
Note that unlike the case of support recovery where the number of measurements had to grow with signal dimension even with $\snr$ of $\log(n)$ here  we see that the number of
measurements does scale with the distortion for moderate signal to noise ratios. This maybe acceptable in cases where either a probability model for the signal set is available.

\begin{thm}
\label{thm:scap_continuous_CS} Consider the output noise model of Equation~(\ref{eq:model_snr}) and the mixture model for $\bX$ as described above. Then,
\begin{itemize}
\item[a.] \textbf{Necessity}: Asymptotically as $n\rightarrow \infty$ if $m \leq \dfrac{n (R_X(d_0)- \frac{\alpha}{2} \log 2)}{0.5 \log (1 + \frac{n}{m}
\alpha \beta^2 \snr)}$ there does not exist any algorithm that recovers the signal to within an average $\ell_2$ distortion of $d_0$.
\item[b.] \textbf{Sufficiency}: Asymptotically as $n\rightarrow \infty$ it is sufficient that
$m \geq \dfrac{n R_X(d_0/2) }{0.5 \log (1 + \frac{n}{m}\frac{d_0 \beta^2 \snr}{2})}$ for the constructive ML estimator of section \ref{sec:peupperbound} to
reliably recover the signal to within an average $\ell_2$ distortion of $d_0$.
\end{itemize}
\end{thm}

\begin{proof}
The proof is similar to the proof of Theorem \ref{thm:scap_continuous}.
\end{proof}

It is easy to see that Corollary \ref{cor:num_meas_discrete} holds true for this case too  with appropriate modifications to the necessary conditions in terms of $R_X(d_0) -
\frac{\alpha}{2} \log 2$ instead of $R_X(d_0)$.
\subsection{Comparison between Worst-Case and Bayesian Setups}
Based on the worst-Case and Bayesian results we can comment on the main differences.  The situation is slightly complicated since we considered two different types of
distortions in these cases. We recall the items {\bf (A)---(D)} listed in the beginning of Section~\ref{sec:average_dist} as a means for comparison. Note that by adopting a
Bayesian setup we no longer need that the minimum singular value of sub-matrices of $\bG$ be uniformly bounded away from zero. This can be attributed to the fact that we are
taking expectation with respect to $\bG$ in Equation~(\ref{eq:PeUpper}). However, note that the number of quantization points $N_{\epsilon}(n,d_0)$ in Theorem~\ref{thm:mincover}
will go to infinity if we insist on nearly exact support recovery. Second, note that the measurements do scale with the distortion-level, larger the admissible distortion,
smaller the number of measurements. This is even more surprising for input noise models since in the worst-case setup we required the number of measurements to scale with signal
dimension. Finally, for signal reconstruction to within a distortion level $d_0$ we only need a constant $\snr$ in contrast to the worst-case setup. However, this issue can be
attributed to the fact that our mean-squared distortion metric is less stringent in comparison to support errors.

\section{Appendix}

\subsection{Proof of Lemma \ref{lem:reduct}}

Consider any arbitrary $\bG$ and $\bN$. Let for each $\bX \in \Xi_{\beta}^{\{k\}}$ denote by $\prob_{\bX}$ the observed distribution of $\bY$ given $\bX$ as induced by the
relation $\bY = \bG \bX + \bN$. We next consider the equivalence class of all sequences with the same support and lump the corresponding class of observation probabilities into
a single composite hypothesis, i.e.,
\begin{align}
[\bX] = \{ \bX' \in \Xi_{\beta}^{\{k\}} \mid \supp(\bX') = \supp(\bX) \}
\end{align}

Each equivalence class bears a one-to-one correspondence with binary valued k-sparse sequences,
\begin{align}
\Xi_{\{0,\beta\}}^{\{k\}} = \{\bX \in \Xi_{\beta}^{\{k\}} \mid X_i = \beta,\,\, i \in \supp(\bX)\}
\end{align}
Our task is to lower bound the worst-case error probability
\begin{align}
\prob_{e|\bG} = \min_{\hat{\bX}} \max_{\bX \in \Xi_{\beta}^{\{k\}}} \prob_\bX ([\hat \bX] \neq [\bX] | \bG)
\end{align}
Now note that,
\begin{align}
\max_{\bX \in \Xi_{\beta}^{\{k\}}} \prob_\bX ([\hat \bX] \neq [\bX]| \bG) \geq \max_{\bX \in \Xi_{\{0,\beta\}}^{\{k\}}} \prob_\bX ([\hat \bX] \neq [\bX] |\bG) = \max_{\bX \in
\Xi_{0,\beta}^{\{k\}}} \prob_\bX (\hat \bX \neq \bX ,\,\,\hat \bX \in \Xi_{\{0,\beta\}} | \bG)
\end{align}
This implies that
\begin{align}
\prob_{e|\bG} & = \min_{\hat \bX \in \Xi_{\beta}^{\{k\}}} \max_{\bX \in \Xi_{\beta}^{\{k\}}} \prob_{\bX}\{\supp(\hat \bX) \neq \supp(\bX) |\bG \} \\ &\geq \min_{\hat
\bX \in \Xi_{\beta}^{\{k\}}}\max_{\bX \in \Xi_{\{0,\beta\}}^{\{k\}}} \prob_\bX (\hat \bX \neq \bX ,\,\,\hat \bX \in \Xi_{\{0,\beta\}}^{\{k\}} | \bG)\\
& = \min_{\hat \bX \in \Xi_{\{0,\beta\}}^{\{k\}}}\max_{\bX \in \Xi_{\{0,\beta\}}^{\{k\}}} \prob_\bX (\hat
\bX \neq \bX ,\,\,\hat \bX \in \Xi_{\{0,\beta\}}^{\{k\}} | \bG)\\
& \geq \min_{\hat \bX \in \Xi_{\{0,\beta\}}^{\{k\}}} \max_{\bX \in \Xi_{\{0,\beta\}}^{'}} \prob_\bX (\hat \bX \neq \bX ,\,\,\hat \bX \in \Xi_{\{0,\beta\}}^{\{k\}}|\bG)
\end{align}
\subsection{Proof of Lemma \ref{lem:superp1}}
Denote by ${\cal E}_{1\omega}$ the error event when a signal from the $\omega$th support set is more likely, i.e.,
\begin{align}
{\cal E}_{1\omega} = \left \{ \bN: \min_{\om \in {\cal I} \bX_{S_{\omega}}^{\min} \geq \beta/2} \| \bY - \bG_{S_{\omega}} \bX_{S_{\omega}}\|^2\leq \min_{ \tilde{\bX} }
\|\bY-\bG_{S_0} \tilde{\bX} \|^2,\,\,\omega\not = 0,\,\, \right \}
\end{align}
In the following we will drop stating the obvious fact that $\om \in {\cal I}$. Now note that,
\begin{align}
{\cal E}_1=\bigcup_{\omega\not = 0}{\cal E}_{1\omega}
\end{align}
We first upperbound ${\cal E}_{1\omega}$ by a more manageable event, namely,
\begin{align}
{\cal F}_{\omega} = \left \{ \bN: \min_{\bX_{S_{0^c,\omega}}^{\min} \geq \beta/2} \min_{\bX_{S_{0,\omega}}}\| \bY - \bG_{S_{0,\omega}} \bX_{S_{0,\omega}} - \bG_{S_{0^c,\omega}}
\bX_{S_{0^c,\omega}}\|^2\leq \min_{ \tilde{\bX} } \|\bY-\bG_{S_0} \tilde{\bX} \|^2,\,\,\omega\not = 0 \right \}
\end{align}
It is clear that,
\begin{align} {\cal E}_{1\omega} \subset {\cal F}_{\omega}\end{align}
This is because the signal on the common support $S_{0,\omega}$ is relaxed to take on any value and not necessarily those that are bounded away from zero by $\beta/2$. We will
now simplify the events in ${\cal F}_{\omega}$ by analytically carrying out the unconstrained minimizations. Recall that $\bY = \bG\bX_0 + \bN$. Let $\bX_{S_0}^{0}$ denote the
true signal $\bX_0$ restricted to its support. Then $\bY = \bG_{S_0}\bX_{S_0}^{0}$. Note that $\bX_{S_0}^{0}$ is composed of $\bX_{S_{0,\om}}^{0}$ corresponding to the overlap
and $\bX_{0,\om^c}^{0}$ corresponding to the misses. We have the following Lemma.
\begin{lem}For $m \geq 2k + 1$
\begin{align}
\label{eq:supp_error_events_A} {\cal F}_{\omega} \subset \tilde {\cal F}_{\omega} = \bigcup_{\bX_{S_{0^c,\omega}}^{\min} \geq \beta/2,\bX_{S_{0,\omega^c}}^{0,\min} \geq \beta }
\left \{ \bN: 2 \bN^T \bPi_1 \bG'\bX' \geq \| \bPi_1 \bG'\bX' \|^2,\,\,
 \right \}
\end{align}
where
\begin{align}
\bPi_1 = (\mathbf{I} - \bG_{S_{0,\om}} (\bG_{S_{0,\om}}^{T}\bG_{S_{0,\om}})^{-1} \bG_{S_{0,\om}}^{T} )
\end{align}
is a projection operator and \begin{align} \bG' = [\bG_{S_{0^c,\om}} \bG_{S_{0,\om^c}}], \bX' =  \begin{bmatrix} - \bX_{S_{0^c,\om}} \\ \bX_{S_{0,\om^c}}^{0} \end{bmatrix} ,\,\,
\bX_{S_{0^c,\om}}^{\min} \geq \beta/2,\,\, \bX_{S_{0,\om^c}}^{0,\min} \geq \beta
\end{align}
\end{lem}
\begin{proof}
Consider the error region,
\begin{align}
{\cal F}_{\omega} = \left \{ \bN: \min_{\bX_{S_{0^c,\omega}}^{\min} \geq \beta/2} \min_{\bX_{S_{0,\omega}}}\| \bY - \bG_{S_{0,\omega}} \bX_{S_{0,\omega}} - \bG_{S_{0^c,\omega}}
\bX_{S_{0^c,\omega}}\|^2\leq \min_{ \tilde{\bX} } \|\bY-\bG_{S_0} \tilde{\bX} \|^2,\,\,\omega\not = 0 \right \}
\end{align}
Fixing $\bX_{S_{0^c,\om}}$ we perform the inner minimization first on the L.H.S in the above equation. It can be shown that the inner minimum is achieved at,
\begin{align}
\bX_{S_{0,\om}}^{0} - \bX_{S_{0,\om}} = -(\bG_{S_{0,\om}}^{T} \bG_{S_{0,\om}})^{-1} \bG_{S_{0,\om}}^{T}( \bN + \bG_{S_{0,\om^c}} \bX_{S_{0,\om^c}}^{0} - \bG_{S_{0^c,\om}}
\bX_{S_{0^c,\om}})
\end{align}
Also the unconstrained minimum on the R.H.S. is given by,
\begin{align}
\min_{\tilde \bX}||(\bY - \bG_{S_0}\tilde \bX)||^2 = \bN^T \bPi_0  \bN
\end{align}
where
\begin{align}
\bPi_0 = (\mathbf{I} - \bG_{S_{0}} (\bG_{S_{0}}^{T}\bG_{S_{0}})^{-1} \bG_{S_{0}}^{T} )
\end{align}
is a projection operator. Substituting these results in the expression for ${\cal F}_{\om}$ we obtain,
\begin{align}
\label{eq:supp_error_events} {\cal F}_{\om} =  & \left\{ \bN: \min_{\bX_{S_{0^c,\omega}}^{\min} \geq \beta/2}  (\bG'\bX')^T \bPi_1\bG'\bX' - 2 \bN^T \bPi_1 \bG'\bX' +
\bN^T(\bPi_0- \bPi_1 )\bN \leq 0 \right\}
\end{align}
A simple application of the matrix lemma shows that $(\bPi_0 - \bPi_1)$ is a positive semi-definite matrix. This implies $\bN^T(\bPi_0 - \bPi_1 )\bN \geq 0$, $\forall \bN$.
Ignoring this non-negative term can only increase the probability of error. Therefore ignoring this term we obtain,
\begin{align}
\label{eq:supp_error_events1} {\cal F}_{\om} =  & \left\{ \bN: \min_{\bX_{S_{0,\omega^c}}^{\min} \geq \beta/2}  (\bG'\bX')^T \bPi_1\bG'\bX' - 2 \bN^T \bPi_1 \bG'\bX'  \leq 0
\right\}\\
& \subseteq \bigcup_{\bX_{S_{0^c,\omega}}^{\min} \geq \beta/2,\bX_{S_{0,\omega^c}}^{0,\min} \geq \beta }\left\{ \bN:  (\bG'\bX')^T \bPi_1\bG'\bX' - 2 \bN^T \bPi_1 \bG'\bX' \leq
0 \right\} \\
& = \bigcup_{\bX_{S_{0^c,\omega}}^{\min} \geq \beta/2,\bX_{S_{0,\omega^c}}^{0,\min} \geq \beta }\left\{ \bN:  2 \bN^T \bPi_1 \bG'\bX' \geq \|\bPi_1\bG'\bX'\|^2 \right\} =
{\tilde{\cal F}}_{\om}
\end{align}
where the last equality follows from the fact that $\bPi_1$ is a projection. Now note that if any column of $\bG'$ falls into the null space of $\bG_{S_{0,\om}}$ then
probability of the event ${\cal F}_{\om}$ is $1$ and therefore the probability of error is $1$ in the worst case. This will not happen as long as $\bG$ is full rank and $m \geq
2k +1$.
\end{proof}
We now have the following Lemma.
\begin{lem}
\label{lem:superp}
\begin{align}
\tilde{\cal F}_{\om} \subseteq {\cal L}_{\om} = \bigcup_{j=1}^L \left\{ \bN: 2\bN^T \bg_{j}^{'} X' \geq \sigma_{\min}((\bG')^T\bG') X'^2 , |X'| =
\beta/2\right\}
\end{align}
where $L = |S_{0^c,\om} \cup S_{0,\om^c} |$ is the total number of location errors and, $\sigma_{\min}((\bG')^T\bG')$ is the minimum singular value of the matrix $(\bG')^T\bG'$.
\end{lem}
\begin{proof}
Let $\tilde \bG = \bPi_1 \bG'$. Then note that for any $\bX'$,
\begin{align}
\left\{\tilde{\bN}: 2 \bN^T\tilde{\bG} \bX' \geq \| \tilde{\bG} \ \bX'\|^2 \right\} \subseteq  \left\{\bN: 2 \bN^T\tilde{\bG} \bX' \geq \sigma_{\min}(\tilde{\bG}^T \tilde{\bG})
\| \bX'\|^2 \right\}
\end{align}
Now note that $$\tilde{\bG} \bX' = \sum_{j=1}^{L}\tilde \bg_{j}X'_j$$ where $\tilde \bg_{j}$ is the $j$-th column of the matrix $\tilde{\bG}$ and $\bX' = [
X'_{1},\ldots,X'_{j},\ldots, X'_{L}]^T$. Note also that $$\|\bX'\|^2 = \sum_j |X'_{j}|^2$$  By a simple \emph{superposition} of events this implies that
\begin{align}
\tilde{\cal F}_{\om}  & \subseteq  \bigcup_{\bX_{S_{0^c,\omega}}^{\min} \geq \beta/2,\bX_{S_{0,\omega^c}}^{0,\min} \geq \beta }
\bigcup_{j=1}^{L} \left\{{\bN}: 2
\bN^T\tilde{\bg}_j X'_j \geq \sigma_{\min}(\tilde{\bG}^T \tilde{\bG}) |X'_j|^2 \right\} \nonumber \\
& \subseteq  \bigcup_{\bX_{S_{0^c,\omega}}^{\min} \geq \beta/2,\bX_{S_{0,\omega^c}}^{0,\min} \geq \beta } \bigcup_{j=1}^L \left\{\bN: 2
\bN^T\tilde{\bg}_j X'_j \geq \sigma_{\min}(\tilde{\bG}^T
\tilde{\bG}) | X'_j|^2 \right\}\\
& \subseteq \bigcup_{j=1}^L \left\{\bN: 2 \bN^T\tilde{\bg}_j X' \geq \sigma_{\min}(\tilde{\bG}^T \tilde{\bG}) |X'|^2 : |X'| = \beta/2 \right\}
\end{align}
where the last inequality follows from the fact all the events with $X' \geq \beta/2$ are contained in the event $X' = \beta/2$. Now note that since $\bPi_1$ is a projection and
$m \geq 2k+1$ and $L \leq 2k$ it implies $\sigma_{min}(\tilde{\bG}^T\tilde\bG) = \sigma_{min}((\bG')^T\bG')$. This implies that,
\begin{align}
\tilde{\cal F}_{\om} & \subseteq \bigcup_{j=1}^L \left\{\bN: 2 \bN^T\tilde{\bg}_j X' \geq
\sigma_{\min}(\tilde{\bG}^T \tilde{\bG}) | X'|^2 : |X'| = \beta/2 \right\}\\
& = \bigcup_{X' = \pm \beta/2} \bigcup_{j=1}^L \left\{\bN: 2 \bN^T \bg'_j X' \geq \sigma_{\bG,\min}((\bG')^T \bG') | X'|^2\right\}
\end{align}
 Since $L \leq 2k \leq n$ and $\left\{\bg'_1,..,\bg'_j,...,\bg'_L\right\} = \left\{\bg_i: i
\in S_{0^c,\omega}\cup S_{0,\omega^c}\right\} \subseteq \left\{\bg_1,...,\bg_n\right\}$,
\begin{align}
{\cal F}_{\om} &\subseteq \bigcup_{X' = \pm \beta/2} \bigcup_{j=1}^n \left\{\bN: 2 \bN^T \bg_j X' \geq \sigma_{\bG,\min} | X' |^2\right\}\\
& = {\cal L}_{\omega}
\end{align}
where $\sigma_{\bG,\min} = \min_{S: |S| \leq 2k} \sigma_{\min}(\bG_{S}^{T} \bG_{S})$.
\end{proof}
The result then follows by noting that,
\begin{align}
{\cal E}_1 = \bigcup_{\om} {\cal E}_{1\om} \subseteq \bigcup_{\om} {\cal F}_{\om} \subseteq \bigcup_{\om} {\cal L}_{\om}
\end{align}
and replacing the notation $X'$ by $X$.

\section{Proof of Lemma \ref{lem:E1}}
From Lemma \ref{lem:superp1} we have ,
\begin{align*}
\prob({\cal E}_1) & \leq  \bigcup_{X = \pm \beta/2} \bigcup_{j=1}^n \left\{\bN: 2 \bN^T \bg_j X \geq \sigma_{\bG,\min}\sqrt{\snr}| X|^2\right\}
 = \bigcup_{X = \pm \beta/2} \bigcup_{j=1}^n \left\{\bW: 2 \bW^T \bsg^{1/2} \bg_j X \geq \sigma_{\bG,\min} \sqrt{\snr}| X|^2\right\} \\
 &=\bigcup_{X = \pm \beta/2} \bigcup_{j=1}^n \left\{w: 2 w X \geq \sqrt{\snr} \sigma_{\bG,\min} \frac{X^2}{\sqrt{\bg_{j}^{'}\bsg \bg_j}}\right\}
\end{align*}
Note that $\bW$ is IID normally distributed Gaussian vector and we let $w= \frac{\bW^T \bsg^{1/2}\bg_j}{\sqrt{\bg_{j}^{T}\bsg \bg_j}}$.
%\begin{align}
%\left\{w: 2 w X \geq \sqrt{\snr} \sigma_{\bG,\min} \frac{X^2}{\sqrt{\|\bg_{j}^{T}\bsg \bg_j\|}}\right\} %= \left\{w: 2 w X \geq \sqrt{\snr} \sigma_{\bG,\min}
%\frac{X^2}{\frac{\sqrt{\|\bg_{j}^{T}\bsg \bg_j\|}}{\|\bg_j\|}}\right\}
%\end{align}
Next noting that $\|\bg_j\| = 1$ $\forall j$ we have,
\begin{align}
\label{eq:E1_1} \left\{w: 2 w X \geq \sqrt{\snr} \sigma_{\bG,\min} \frac{X^2}{\sqrt{\|\bg_{j}^{T}\bsg \bg_j\|}}\right\} & \subseteq  \left\{w: 2 w
X \geq \sqrt{\snr} \sigma_{\bG,\min}  \sqrt{\frac{1}{\lambda_{\max}(\bsg)}}X^2\right\} \\
& = \left\{w: 2 w X \geq \sqrt{\snr} \sigma_{\bG,\min} \sqrt{\lambda_{\min}(\bsg^{-1})} X^2\right\}
\end{align}
We now apply the union bound over all the possible $2n$ error events corresponding to each $j \in \{1,2,...,n\}$ and $X = \pm \beta/2$ and obtain,
\begin{align}
\prob({\cal E}_1) & \leq  \prob\left\{w:  w \geq \sqrt{\lambda_{\min}(\bsg^{-1})}\sigma_{\bG,min}\frac{\sqrt{\snr}\beta}{4} \right\} \exp(\log 2n)\\
& = \frac{1}{\sqrt{2\pi}}\int_{\sqrt{\lambda_{\min}(\bsg^{-1})} \sigma_{\bG,\min}\frac{\sqrt{\snr}\beta}{4} }^{\infty} \exp(-y^2/2) dy\,\cdot \exp(\log 2n) \\
\label{eq:E1_end} &\leq \exp \left\{- \lambda_{\min}(\bsg^{-1})\sigma_{\bG,\min}^{2}\frac{ \beta^2 \snr}{ 32}\right\} \exp \left\{\log 2n\right\}
\end{align}
Note that the probability is only taken over the noise $\bW$ ($\bN$) as $\bG$ is given and is fixed. Here we have used the approximation ${\cal Q}(x) \leq \exp(-x^2/2)$ for the
standard error function defined as ${\cal Q}(x) = \frac{1}{\sqrt{2\pi}} \int_{x}^{\infty} \exp(-x^2/2) dx$.

\section{Proof of Lemma \ref{lem:E2}}
For \emph{any} $\bX_0$ supported on the submatrix $\bG_{S_0}$ the probability of the error event ${\cal E}_2$ is given by,
\begin{align}
\prob({\cal E}_2) = \prob \left\{\bN: \|(\bG_{S_0}^{T}\bG_{S_0})^{-1}\bG_{S_0}^{T} \bN\|_\infty\geq \sqrt{\snr}\beta/2\right\}
\end{align}
To this end let $\bG_{S_0} = \bU \bsg_{S_0} \bV^*, \bU \in \Complex^{m\times k}, \bV^* \in \Complex^{k \times m}$. Then $ (\bG_{S_0}^{T}\bG_{S_0})^{-1}\bG_{S_0}^{T} = \bU
\bsg_{S_0}^{-1} \bV^{*}$. Then let $\tilde \bN = \bV^* \bN$. Then since $\bV$ is orthonormal matrix $\tilde \bN$ has the same distribution as that of $\bN$. Now note that if
$\bN \sim {\cal N}(0,\bsg)$, then
\begin{align}
\prob\left\{\| \bU \bsg_{S_0}^{-1} \tilde \bN\|_{\infty} \geq \snr \beta/2\right\}
&\leq \sum_{i=1}^{m}  \prob\left\{\| (\bU \bsg_{S_0}^{-1} \bsg^{1/2} \bW)_i\| \geq \frac{ \snr \beta}{2 }\right\} \\
& \overset{(a)}{\leq} 2 m
\exp\left\{-\frac{\snr \beta^2}{8}\lambda_{\min}(\bsg^{-1})\sigma_{\bG_{S_0},\min}^{2}\right\}\\
& \overset{(b)}{\leq} exp\left\{-\frac{\snr \beta^2 \lambda_{\min}(\bsg^{-1})\sigma_{\bG,\min}^{2}}{8} + \log 2 n\right\}
\end{align}
where (a) follows from the following facts applied in succession- (1) Maximum variance among the noise components $(\bU \bsg_{S_0}^{-1} \bsg^{1/2} \bW)_i$ is given by
$\sigma_{\bG_{S_0,\min}}^{-1} \lambda_{\max}(\bsg^{1/2})$ and $\lambda_{\max}(\bsg^{1/2}) = \lambda_{\min}(\bsg^{-1/2})$; (2) ${\cal Q}(x) \leq e^{-x^2/2}$ for the standard
error function defined as ${\cal Q}(x) = \frac{1}{\sqrt{2\pi}} \int_{x}^{\infty} \exp(-x^2/2) dx$. (b) follows from the fact that $m \leq n$ and $\sigma_{\bG,\min} \leq
\sigma_{\bG_{S_0},\min}$.
\subsection{Proof of Theorem \ref{thm:dist_supp}}
We follow along the lines of the proof for the deterministic case presented in Section~\ref{sec:suffD}. Basically we modify Lemma \ref{lem:superp1}. We follow the same steps till Lemma \ref{lem:superp}. Then following similar algebraic steps as used in Lemma
\ref{lem:superp} it turns out that the support error events with Hamming distortion $\geq 2kd_0 + 1$ are almost contained in the union of support error events with Hamming
distortion $kd_0 \leq d_H \leq 2 kd_0 $. Then in this case the upper bound in Proposition \ref{prp:Pe_general} is modified to,
\begin{align}
\prob_{e|\bG} \leq 2 \exp\left\{- \bsg^{-1}) \sigma_{\bG,\min}^{2}\frac{ \beta^2 k d_0 \snr}{ 32}\right\} e^{2 kd_0 H_2(2kd_0/n)}
\end{align}
The result for Gaussian $\bG$ is then identical to the development in Section~\ref{sec:suffG}.
\subsection{Proof of lemma \ref{lem:lowerbound}}
Let $X^{n} = \left\{X_1,\ldots,X_n\right\}$ be an IID sequence where each variable $X_i$ is distributed according to a distribution $P_{X}$ defined on the alphabet ${\cal X}$.
Denote $P_{X^n} \triangleq (P_{X})^n$ the n-dimensional distribution induced by $P_{X}$. Let the space ${\cal X}^{n}$ be equipped with a distance measure $d(.,.)$ with the
distance in $n$ dimensions given by $d(X^n,Z^{n}) = \sum_{k=1}^{n} d(X_k,Z_k)$ for $X^n, Z^n \in {\cal X}^n$. For this setting we have the following Theorem taken from
\cite{Kontoyiannis01}.

\begin{thm}
\label{thm:mincover} Given $\epsilon >0$, there exist a set of points $\left\{Z_{1}^{n},...,Z_{N_{\epsilon}(n,d_0)}^{n}\right\} \subset {\cal X}^n$ such that,
\begin{eqnarray} \label{eq:cover1} P_{X^n}\left(\bigcup_{i=1}^{N_{\epsilon}(n,d_0)} {\cal B}_{i}\right) \geq 1 - \epsilon
\end{eqnarray}
where ${\cal B}_{i} \triangleq \left\{ X^n : \frac{1}{n} d(X^n,Z_{i}^{n}) \leq d_0 \right\}$ with the property that $\frac{1}{n} \log N_{\epsilon}(n,d_0) \leq R_{X}(d_0) +
\epsilon$. This implies that for all $X^n$, $\exists$ a mapping $f(X^n): X^n \rightarrow Z_{i}^{n}\,\, s.t. \,\, \prob\left(\frac{1}{n} d(X^n,Z_{i}^{n}) \leq d_0\right) \geq 1 -
\epsilon$
\end{thm}

Now we are given that there is an algorithm $\hat{X}^n(\bY)$ that produces an estimate of $X^n$ given the observation $\bY$. To this end define an error event on the algorithm
as follows,
\[ E_n =
\left\{\begin{array}{l}
 1  \,\,\mbox{if} \,\,\, \frac{1}{n}d(X^n,\hat{X}^n(\bY)) \geq d_0 \\
 0  \,\, \mbox{otherwise}
  \end{array}\right. \]
Now, consider the following expansion,
\begin{align*}
H(f(X^n),E_n,|\bY) & =  H(f(X^n)|\bY) + H(E_n, A_n | f(X^n),\bY) \\
& =  H(E_n|\bY) + H(f(X^n)|E_n,\bY)
\end{align*}
This implies that
\begin{align*}
H(f(X^n)|\bY) \leq H(E_n) +  H(f(X^n)|E_n,\bY)
\end{align*}
Note that since $H(E_n) \leq 1$
\begin{align}
H(f(X^n)|\bY) \leq 1 + \prob_{e} H(f(X^n)|\bY,E_n = 1 ) + (1 -\prob_{e})H(f(X^n)|\bY,E_n = 0)
\end{align}
Note that by construction $H(f(X^n)|\bY,E_n = 1 ) \leq \log N_{\epsilon}(n,d_0)$ and  $(1 -\prob_{e})H(f(X^n)|\bY,E_n = 0 ) \leq (1- \prob_{e}^{n}) \log \left( |{\cal S}|
\right)$ where ${\cal S}$ is the set given by,
\[ {\cal S} = \left\{ i : d_{set}\left({{\cal B}_{f(X^n)}, \cal B}_{i}\right)
\leq n d_0 \right\} \] where $d_{set}(S_1,S_2) = \min_{s \in S_1, s' \in S_2}d_{n}(s,s')$ is the set distance between two sets. Now note that $H(f(X^n)|\bY) = H(f(X^n)) -
\ind(f(X^n);\bY) \geq H(f(X^n)) - \ind(X^n;\bY)$ where the second inequality follows from data processing inequality over the Markov chain $f(X^n) \leftrightarrow X^n
\leftrightarrow \bY$. Thus we have,
\begin{align}
\prob_{e} & \geq   \dfrac{H(f(X^n)) - \log |{\cal S}| - \ind(X^n;\bY) - 1}{\log N_{\epsilon}(n,d_0) - \log |{\cal S}|} \\
& \geq \dfrac{\ind(f(X^n);X^n) - \log |{\cal S}| - \ind(X^n;\bY) - 1}{n R_X(d_0) + \epsilon}
\end{align}
The proof then follows by noting that by definition of the rate distortion function $\ind(f(X^n);X^n) \geq n R_X(d_0)$ (see \cite{Cover}) and by identifying $K(n,d_0) =
\frac{1}{n} \log |{\cal S}|$.

\subsection{Proof of lemma \ref{lem:discrete_LB} }

\begin{proof}
Define the error event,
\[ E =
\left\{\begin{array}{l}
 1  \,\,\mbox{if} \,\,\, \frac{1}{n}d_{H}(X^n,\hat{X}^n(\bY)) \geq d_0 \\
 0  \,\, \mbox{otherwise}
  \end{array}\right. \]
Expanding $H(X^n,E|\bY)$ in two different ways we get that,
\[ H(X^n|\bY) \leq 1 + n \prob_e \log (|{\cal X}|) + (1 - \prob_e) H(X^n| E =
0,\bY) \] Now the term
\begin{align} ( 1 - \prob_e) H(X^n | E = 0, \bY) & \leq (1 - \prob_e)
\log \sum_{j = 0}^{nd_0 - 1} \binom{n}{d_0 n - j}(|{\cal X}| -1)^{nd_0 - j}  \\
& \leq (1 - \prob_e) \log nd_0 \binom{n}{d_0 n -
1}(|{\cal X}| -1)^{nd_0}\\
& \leq n (1- \prob_e) \left(H_2(d_0) + d_0 \log (|{\cal X}| -1)  + \frac{\log nd_0}{n} \right)
\end{align}
where the second inequality follows from the fact that $d_0 \leq 1/2$ and $ \binom{n}{d_0 n - j}(|{\cal X}| -1)^{nd_0 - j}$ is a decreasing function in $j$ for $d_0 \leq 1/2$.
Then we have for the lower bound on the probability of error that,
\[ \prob_e \geq \frac{ H(X^n|\bY) - n \left(H_2(d_0) + d_0 \log (|{\cal X}|
 -1) +  \frac{\log nd_0}{n}\right)  - 1}{n \log(|{\cal X}|) - n
\left(H_2(d_0) + d_0 \log (|{\cal X}|-1) + \frac{\log nd_0}{n} \right) }
\]
Since $H(X^n|\bY) = H(X^n) - \ind(X^n;\bY)$ we have
\[ \prob_e \geq \frac{n\left(H(X)  - H_2(d_0) - d_0 \log (|{\cal X}|
-1) - \frac{\log nd_0}{n}\right)   - \ind(X^n;\bY) - 1}{n \log(|{\cal X}|) - n \left(H_2(d_0) + d_0 \log (|{\cal X}|-1) +  \frac{\log nd_0}{n} \right) } \] It is known that
$R_{X}(d_0) \geq  H(X)  - H_2(d_0) - d_0 \log (|{\cal X}| -1) $, with equality iff \[ d_0 \leq (|{\cal X}|-1)\min_{x \in {\cal X}} P_{X}(x) \] see e.g., \cite{Csiszar}. Thus for
values of distortion $d_0$,
\begin{align}
d_0 \leq \min \left\{ 1/2, (|{\cal X}|-1)\min_{x \in {\cal X}} P_{X}(x) \right\}
\end{align}
we have for all $n$,
\[ \prob_e \geq \frac{nR_{X}(d_0) - \ind(X^n;\bY) - 1 - \log
nd_0}{n \log(|{\cal X}|) - n \left(H_2(d_0) + d_0 \log (|{\cal X}|-1) + \frac{\log nd_0}{n} \right)}  \]
\end{proof}

\subsection{Rate distortion function for the mixture Gaussian source
under squared distortion measure} \label{subsec:ratedist_gauss}

It has been shown in \cite{zamirIT02} that the rate distortion function for a mixture of two Gaussian sources with variances given by $\sigma_1$ with mixture ratio $\alpha$ and
$\sigma_0$ with mixture ratio $1- \alpha$, is given by
\[ \begin{array} {l} R_{mix}(D) = \\ \left\{ \begin{array}{l}
H_2(\alpha) + \frac{(1 - \alpha)}{2} \log(\frac{\sigma_{0}^{2}}{D}) + \frac{\alpha}{2}
\log (\frac{\sigma_{1}^{2}}{D}) \,\, \mbox{if} \,\, D < \sigma_{0}^{2} \\
H_2(\alpha) + \frac{\alpha}{2} \log(\frac{\alpha \sigma_{1}^{2}}{D - (1 - \alpha)\sigma_{0}^{2}}) \,\, \mbox{if} \,\, \sigma_{0}^{2} < D \leq (1 - \alpha)\sigma_{0}^{2} + \alpha
\sigma_{1}^{2} \end{array} \right. \end{array}\] For a strict sparsity model we have $\sigma_{0}^2 \rightarrow 0$ we have
\[ R_{mix}(D) = \begin{array}{l} H_2(\alpha) + \frac{\alpha}{2}
\log(\frac{\alpha \sigma_{1}^{2}}{D} ) \,\, \mbox{if} \,\, 0 < D \leq \alpha \sigma_{1}^{2} \end{array}  \]

\bibliographystyle{IEEEtran}
\bibliography{Thesis_new_Scap}

\end{document}